\makeatletter
\declare@file@substitution{revtex4-1.cls}{revtex4-2.cls}
\makeatother
\documentclass[twocolumn]{aastex631}
\usepackage{graphicx} 
\usepackage[caption=false]{subfig}
\usepackage{apjfonts}
\usepackage{amsmath, amstext}
\usepackage{hyperref}
\usepackage{listings}
\usepackage{savesym}
\savesymbol{tablenum}
\usepackage{siunitx}
\restoresymbol{SIX}{tablenum}

\usepackage{ulem} 
\usepackage{xspace}
\usepackage{placeins, listings, booktabs}
\usepackage{tikz}
\usepackage{lipsum}
\usepackage{colortbl}
\usepackage{tikz}
\usepackage{circuitikz}

\usetikzlibrary{math} 
\tikzmath{\s = 0.3;
	\A= 0; \B =1*\s; \C=2*\s;\D=3*\s;\E=4*\s;\F=5*\s;	
	\Z1=0; \Z2=1*\s; \Z3=2*\s; \Z4=3*\s; \Z5=4*\s; \Z6 = 5*\s;
	\Y1=0; \Y2=-1*\s; \Y3=-2*\s; \Y4=-3*\s; \Y5=-4*\s; \Y6 = -5*\s;
		}

\newcommand{\Planck}{\textit{Planck}\xspace}
\newcommand{\WMAP}{\textit{WMAP}\xspace}

\usepackage{etoolbox}
\makeatletter
\patchcmd\linenumberpar{\@LN@parpgbrk}{\penalty\@LN@parpgpen\relax}{}{}
\makeatother

\date{\today}

\begin{document}

\title{Design and Performance of 30/40\,GHz Diplexed Focal Plane for BICEP Array} 

\shorttitle{Design and Performance of a 30/40\,GHz Diplexed Focal Plane for BICEP Array}

\newcommand\Princeton{ Princeton University, Princeton, NJ 08544, USA} 
\newcommand\Caltech{California Institute of Technology, Pasadena,
CA 91125, USA}
\newcommand\JPL{Jet Propulsion Laboratory, Pasadena, CA 91109 USA} 

\author{Corwin~Shiu}
\affiliation{\Princeton} 

\author{Ahmed~Soliman}
\affiliation{\Caltech}
\affiliation{\JPL}

\author{Roger~O'Brient}
\affiliation{\Caltech}
\affiliation{\JPL}

\author{Bryan~Steinbach}
\affiliation{\Caltech}

\author{James~J.~Bock}
\affiliation{\Caltech}
\affiliation{\JPL}
\author{Clifford~F.~Frez}
\affiliation{\JPL}
\author{William~C.~Jones}
\affiliation{\Princeton}
\author{Krikor.~G.~Megerian}
\affiliation{\JPL}
\author{Lorenzo~Moncelsi}
\affiliation{\Caltech}
\author{Alessandro~Schillaci}
\affiliation{\Caltech}
\author{Anthony~D.~Turner}
\affiliation{\JPL} 
\author{Alexis~C.~Weber} 
\affiliation{\JPL}
\author{Cheng~Zhang}
\affiliation{\Caltech}
\author{Silvia~Zhang}
\affiliation{\Caltech}

\begin{abstract}

We demonstrate a wide-band diplexed focal plane suitable for observing low-frequency foregrounds that are important for cosmic microwave background polarimetry.  
The antenna elements are composed of slotted bowtie antennas with 60\% bandwidth that can be partitioned into two bands. 
Each pixel is composed of two interleaved 12$\times$12 pairs of linearly polarized antenna elements forming a phased array, designed to synthesize a symmetric beam with no need for focusing optics. The signal from each antenna element is captured in-phase and uniformly weighted by a microstrip summing tree. 
The antenna signal is diplexed into two bands through the use of two complementary, six-pole Butterworth filters. This filter architecture ensures a contiguous impedance match at all frequencies, and thereby achieves minimal reflection loss between both bands. 
Subsequently, out-of-band rejection is increased with a bandpass filter and the signal is then deposited on a transition-edge sensor bolometer island. 
We demonstrate the performance of this focal plane with two distinct bands, 30 and 40\,GHz, each with a bandwidth of $\sim 20$ and $15\,$GHz, respectively. The unequal bandwidths between the two bands are caused by an unintentional shift in diplexer frequency from its design values. 
The end-to-end optical efficiency of these detectors are relatively modest, at 20-30\%, with an efficiency loss due to an unknown impedance mismatch in the summing tree. Far-field beam maps show good optical characteristics with edge pixels having no more than $\sim$ 5\% ellipticity and $\sim$10-15\% peak-to-peak differences for A-B polarization pairs.

\end{abstract}

\keywords{cosmology: cosmic background radiation --- instrumentation: detectors --- instrumentation: polarimeters}

\section{Introduction} 
 Measurements of the spatial anisotropies of the cosmic microwave background (CMB) provide fundamental tests of cosmological theories. 
 Measurements of a cosmological, degree-scale, B-mode polarization would provide a constraint on the tensor-to-scalar ratio $r$ and place limits on the energy scale of inflation \citep{Seljak_1997, Kamionkowski_2001}. The BICEP experiment has placed the tightest constraints on $r$ with an upper limit of $r_{0.05} < 0.036$ at 95\% confidence \citep{bicep2018}. 

 These measurements are complicated by the fact that cosmological signals at large angular scales are highly contaminated by astrophysical foregrounds. Thermal emission from spinning dust grains produces a polarized emission that dominates at high frequencies \citep{Finkbeiner_1999, Planck_2018_XI}. Galactic synchrotron emission is emitted by electrons gyrating in magnetic fields and dominates at low frequencies \citep{ginzburg_1969, wmap_nineyear}. CMB polarimetry experiments must have several frequency bands to remove contamination from foreground emission and uncover the underlying cosmological signal \citep{brandt1994}. 

Much remains to be learned about characterizing low-frequency foregrounds' spectral and spatial behavior \citep{planck_2015_XXV}.
\WMAP has observed substantial amounts of polarized foreground emission due to synchrotron radiation, even at high Galactic latitudes \citep{Page_2007}, and this has been confirmed by other experiments \citep{Krachmalnicoff_2018, eimer_2023}.  
\cite{Kogut_2007} has observed a flattening of the synchrotron spectral index closer to the Galactic plane; however, \cite{Choi_2015} has attributed this effect to the spatial correlation of dust and synchrotron. 

A cross-correlation analysis of BICEP 95 and 150\,GHz data up to 2018, when combined with publicly available \WMAP K+Ka bands and \Planck NPIPE 30 and 44\,GHz data, did not yield any statistically significance evidence supporting the detection of synchrotron radiation \citep{bicep2018}.
Therefore, while synchrotron contamination is not a driving source of uncertainty for current upper limits on $r$, as experiments become more sensitive, this foreground will play a more significant role in an unbiased recovery of $r$. 

Furthermore, although synchrotron radiation is the primary source of low-frequency, polarized foregrounds, it may not be the exclusive source of low-frequency contamination. 
Magnetic dipole emissions resulting from thermal fluctuations of ferromagnetic interstellar grains have been proposed \citep{drain_lazarian, draine_hensley}. 
These emissions are expected to primarily occur $\nu \leq 100$\,GHz, but the behavior likely differs whether they are from iron inclusions in dust grains or free-flying iron nanoparticles \citep{Hoang_lazarian}. 
While there is currently no statistically significant detection of polarized anomalous microwave emission (AME) \citep{planck_2015_XXV, Herman_2023}, increasingly precise measurements have the potential to improve the understanding of low-frequency foregrounds.

Advances in mm and sub-mm wave bolometer arrays directly drive improvements in characterizing the millimeter-wave sky. Ground-based CMB experiments have long been photon-noise limited \citep{bicep2_threeyear}, and therefore, the sensitivity of an experiment scales with the square root of the number of detectors on the sky. For a given optical system and fixed focal plane area, the sensitivity would improve by taking advantage of a wide bandwidth antenna partitioned into multiple spectral bands. Various technologies exist to couple electromagnetic radiation onto bolometer arrays ranging from lenslet-coupled broadband sinuous antennas to multichroic horn antennas \citep{roger2013, mcmahon2012}. 
Several experiments have successfully deployed multichroic detector arrays, including ACT \citep{ACT_2016} and SPT \citep{spt_2014}, and future generation experiments are all multichroic \citep{simonsarray_experiment, Kiuchi_2020, Walker_2020}.

 This paper describes the design and performance of a fully lithographed, diplexed, low-frequency focal plane that will be powerful for characterizing synchrotron emission. The design builds upon previous BICEP instruments' successful phased array design \citep{antenna_coupled_tes_bolo}. The phased array configuration gives us several advantages. The detectors are entirely planar and fully lithographed in thin films, which significantly simplifies the fabrication of these devices. The phased array configuration naturally synthesizes the beam without any need for focusing optics. 

Antenna arrays have flexibility in their illumination pattern and can match feedhorn-arrays aperture efficiencies of $\sim 0.70$, at smaller pixel sizes than their feedhorn counterparts \citep{Griffin_2002}. Consequently, pixel densities from antenna arrays can be higher, owing to both geometric tiling of square pixels and an increased number of detectors within the same focal plane area. These theoretical advantages in efficiency arise from the ability for antenna arrays to have greater directivity.
An extended discussion can be found in Appendix \ref{appendix:aperture_eff}, which outlines the theoretical contributions to focal plane mapping speed.

\section{Focal Plane Overview}  
\begin{figure*}
\centering
\begin{minipage}{0.35\textwidth}
\includegraphics[width = \textwidth]{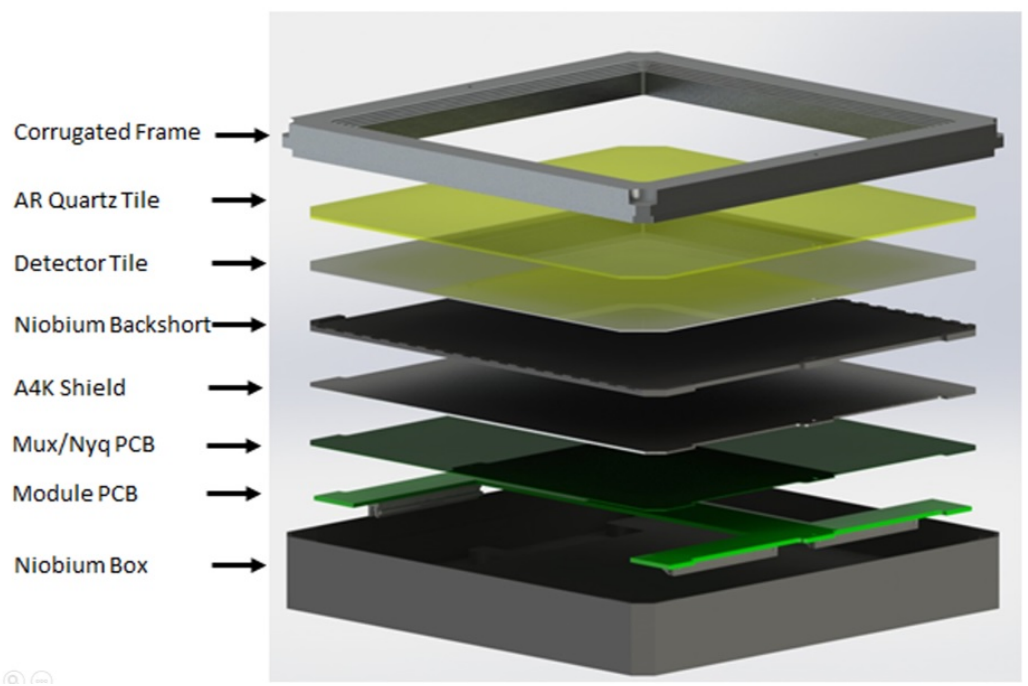}
\end{minipage}
\begin{minipage}{0.5\textwidth}
\includegraphics[width = \textwidth]{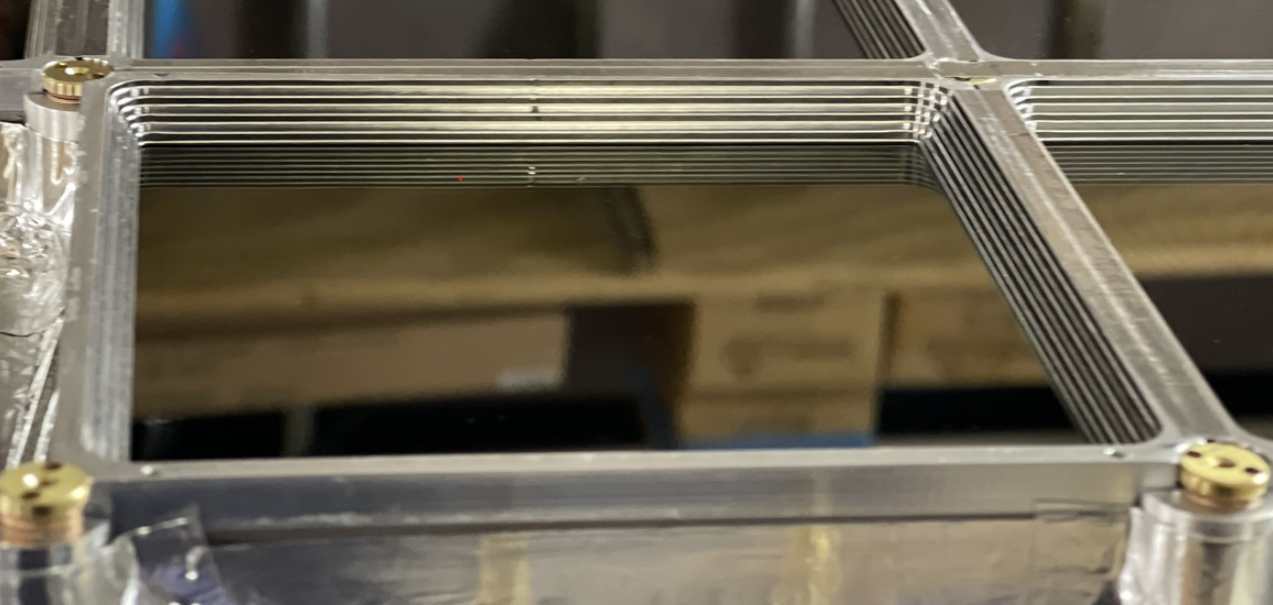}
\end{minipage}
\caption{(Left) A cross-sectional 3D render of the focal plane module. The focal plane module consists of a detector tile hybridized with its readout chain, all housed in a superconducting niobium box and frame for magnetic shielding. (Right) A photograph of the machined corrugated frames designed to minimize undesired electrical interactions between the edge-pixel antenna elements and the structure. The visible tile is the anti-reflection coating, facilitating the backside illumination of the detectors through the silicon.  }
\label{fig:detector_module}
\end{figure*}
\begin{figure*}
\includegraphics[width = 0.23\textwidth]{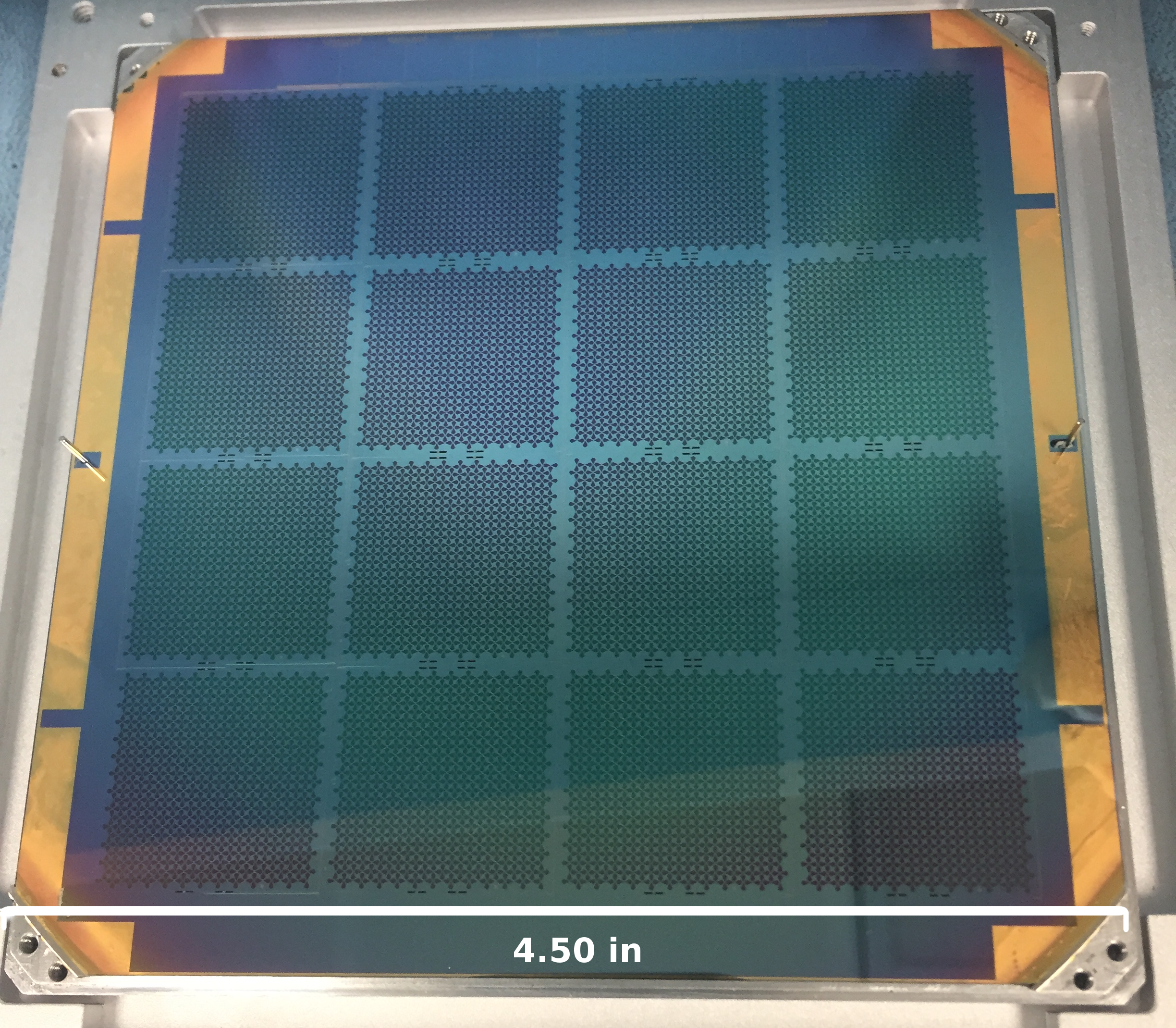}
\includegraphics[width = 0.23\textwidth]{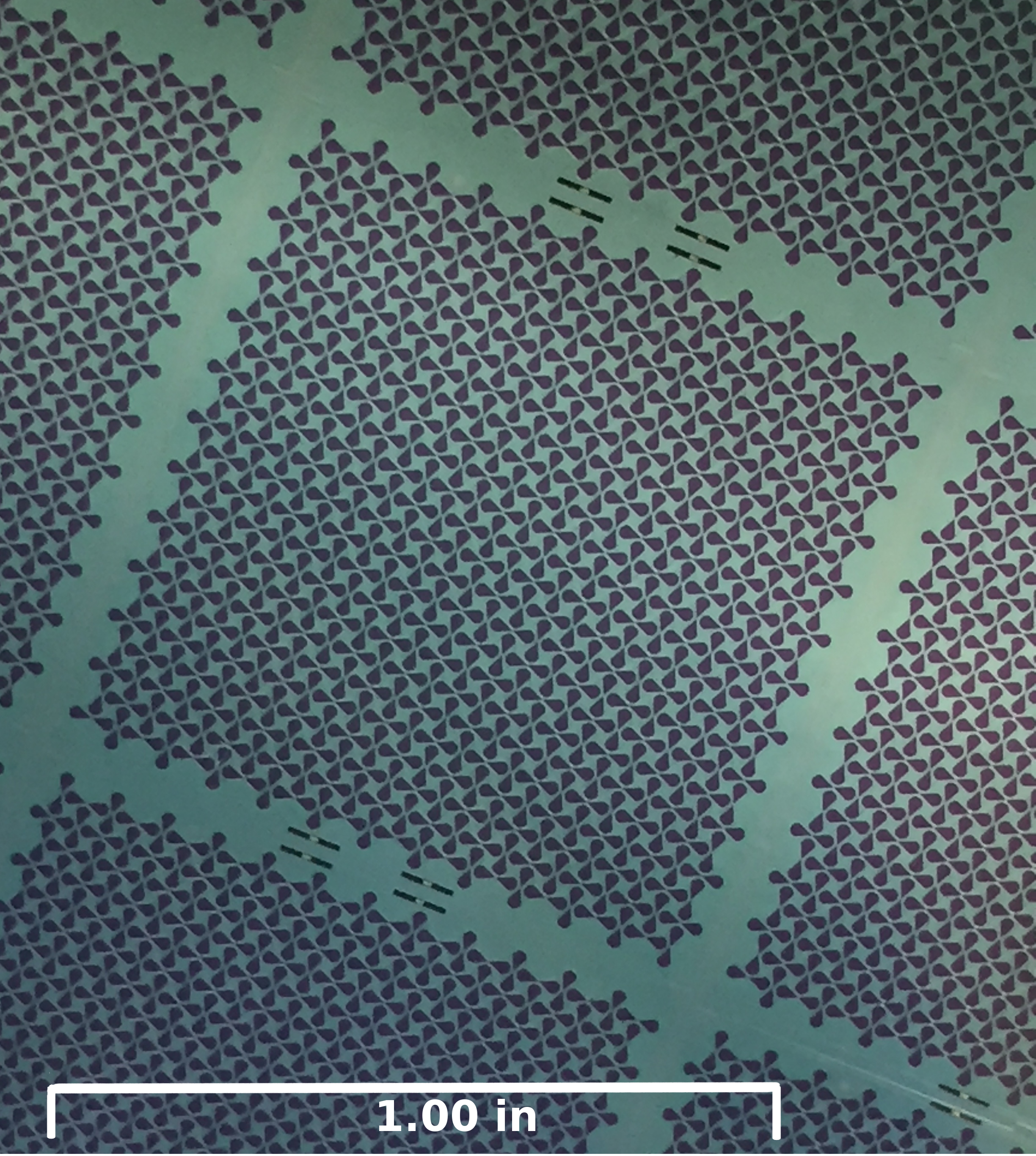} 
\includegraphics[width = 0.45\textwidth]{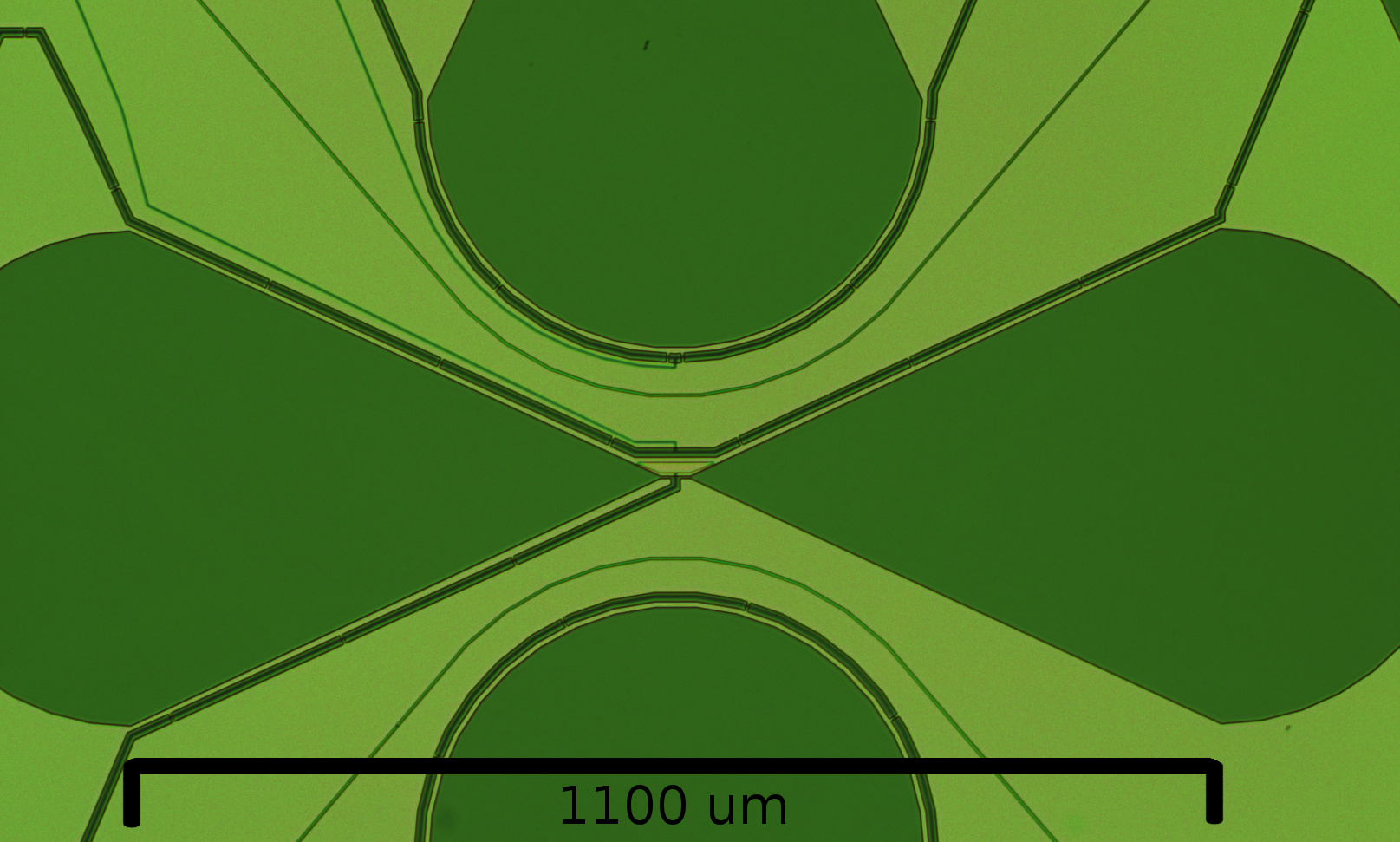}
\caption{Bowtie tile. (a) Photograph of a full tile showing a 4x4 grid of pixels. (b) Zoom in on a pixel. We have a 12x12 pairs of bowtie antennas comprising the antenna array with four bolometers capturing A/B polarization at our two bands. (c) Zoom in on a single antenna element. Dark green indicates the ground plane cutout showing both the sub-antenna and the co-planar waveguide driving the antenna.} 
\label{fig:focalplane}
\end{figure*}

Our detector design is entirely planar and requires no contacting optics such as lenslets or horn antennas. 
Incoming optical power is coupled to polarized planar antenna arrays, the details of which are provided in section \ref{sec:antenna_design}. The power is then directed through a microstrip summing network, as explained in section \ref{sec:microstrip_fed}. 
The power subsequently passes through on-chip band-defining filters, which partition the power by frequency; the design of these filters is elaborated upon in section \ref{sec:filtering}. 
Finally, the energy is dissipated on a bolometer island and detected by a superconducting TES \citep{kent_irwin_TES}, the design of which is explained in section \ref{sec:bolometer}. Variations in the TES current are readout by a time-domain multiplexing system based on SQUIDS \citep{tmux}. 

\subsection{Overview of Detector Wafer}
The Microdevices Laboratory at the Jet Propulsion Laboratory (JPL) fabricates the detector wafers.
Antenna arrays are built on 6\,in, 625\,$\mu m$ thick silicon wafers ($\epsilon_r = 11.8$)
The millimeter wave circuit consists of four distinct layers, described in order from the layer closest to the silicon to the layer farthest away. 
 First, a niobium ground plane film is deposited, and then slotted antenna arrays are patterned through a liftoff process. 
 Subsequently, we grow a 0.3\,$\mu$m SiO$_2$ inner-layer dielectric (ILD) on top. Following that, we define the resistive termination for the bolometers. Finally, in the last film, we pattern the upper niobium conductor, which is responsible for shaping the antenna feed network and in-line filters. 
 Fabrication details can be found in \cite{antenna_coupled_tes_bolo}. 
Figure \ref{fig:focalplane} shows a photograph of the antenna elements. 

Because the antenna arrays are backside illuminated, a fused-quartz ($\epsilon_r$ = 3.9) anti-reflection layer is applied to the bottom of the entire stack, serving as the topmost layer facing the sky. As a result, the antenna elements face a superconducting niobium reflective backshort, positioned at a distance of $\lambda/4$ away.
The entire detector module comprises the quartz anti-reflection wafer, silicon detector array, niobium backshort, amumetal 4K magnetic shielding, and readout PCB cards, all enclosed in a compact niobium frame.
Additional information on the focal plane engineering, measurements, and hybridization can be found in other publications \citep{Ahmed_thesis, schillaci_detectormodule}.

\subsection{Mitigating Polarized Frame Edge-Effects}

Unwanted electromagnetic interactions between the niobium frame and antennas degrade the quality of the antenna beams. Employing a solid metal frame can lead to polarization-dependent deformations of the antenna beams, especially for edge pixels. 
Such deformations can introduce (1) differential beam centers for orthogonal polarizations and result in dipole artifacts or (2) polarized beam ellipticies and result in quadrupole artifacts. These beam systematics leak temperature to polarization and potentially introduce a false B-mode signal \citep{Hu_2003, bicep2_instrument_systematics}. 
Although BICEP has developed a deprojection technique to mitigate low-order beam systematics effectively, controlling higher-order beam effects is challenging \citep{bicep_2016}. 

At lower frequencies, there are far more edge pixels than center pixels. \cite{soliman_3040frame} has devised a novel corrugated frame that minimizes polarized beam-steering. These grooves serve to smooth out impedance discontinuities, effectively reducing the abrupt boundaries caused by the presence of a conductive frame. 
The frame relies on quarter-wave corrugations with quarter-wave pitches. This design ensures that surface waves reflecting off the corrugated frame become out of phase and destructively interfere. 
To account for the wide bandwidth of this antenna, \cite{Ahmed_thesis} designed a doubly corrugated frame with depths tailored for both 30 and 40\,GHz. 
The peak-to-peak polarization subtraction for
the broad-band corrugated frame shows a minimized differential offset ($<10-15\%$). 
The remaining residuals are effectively filtered out using a deprojection technique developed by the BICEP team. 
These results show that the corrugated frame successfully minimizes the differential offset compared to earlier FPU detector designs; \cite{bk15beam} reports a differential offset of $\sim 40\%$. Readers interested in the design specifications of this corrugated frame are referred to a different publication \citep{Ahmed_thesis}.

\subsection{Antenna Design} 
\label{sec:antenna_design} 

\begin{figure*}
\centering
\includegraphics[width = 0.35\textwidth]{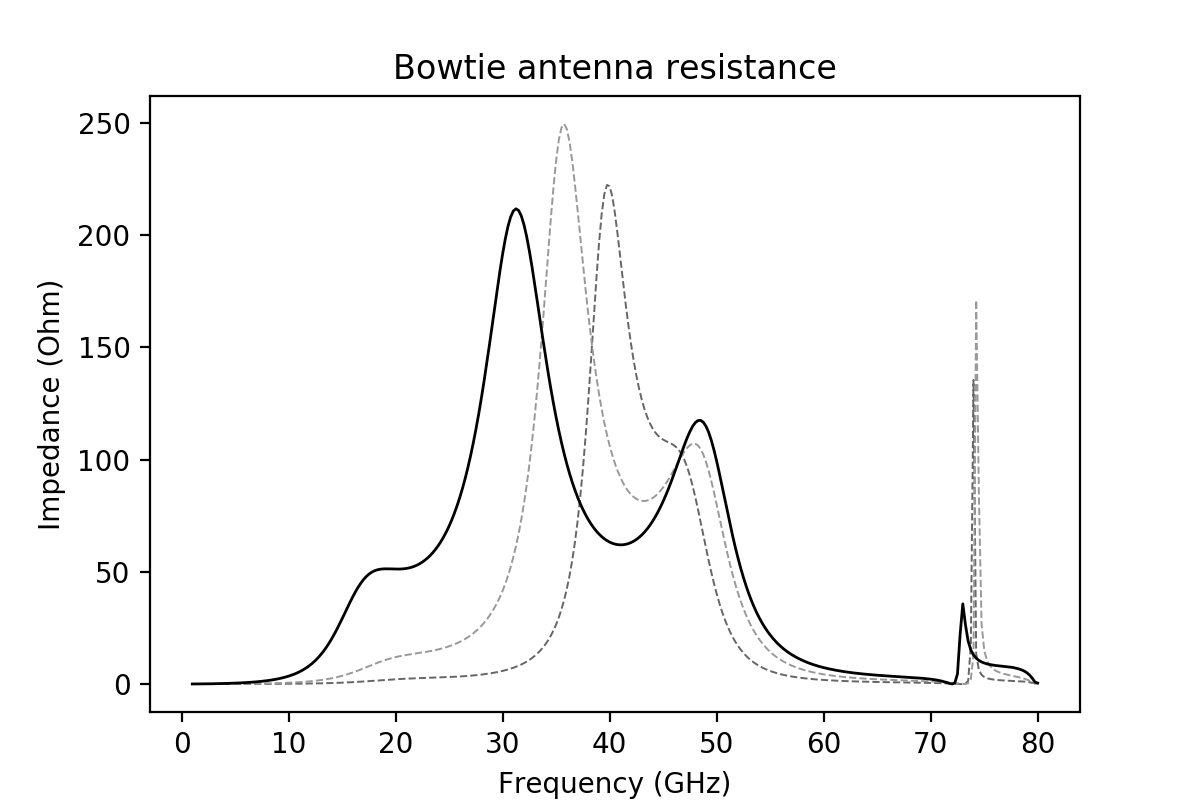}
\includegraphics[width = 0.35\textwidth]{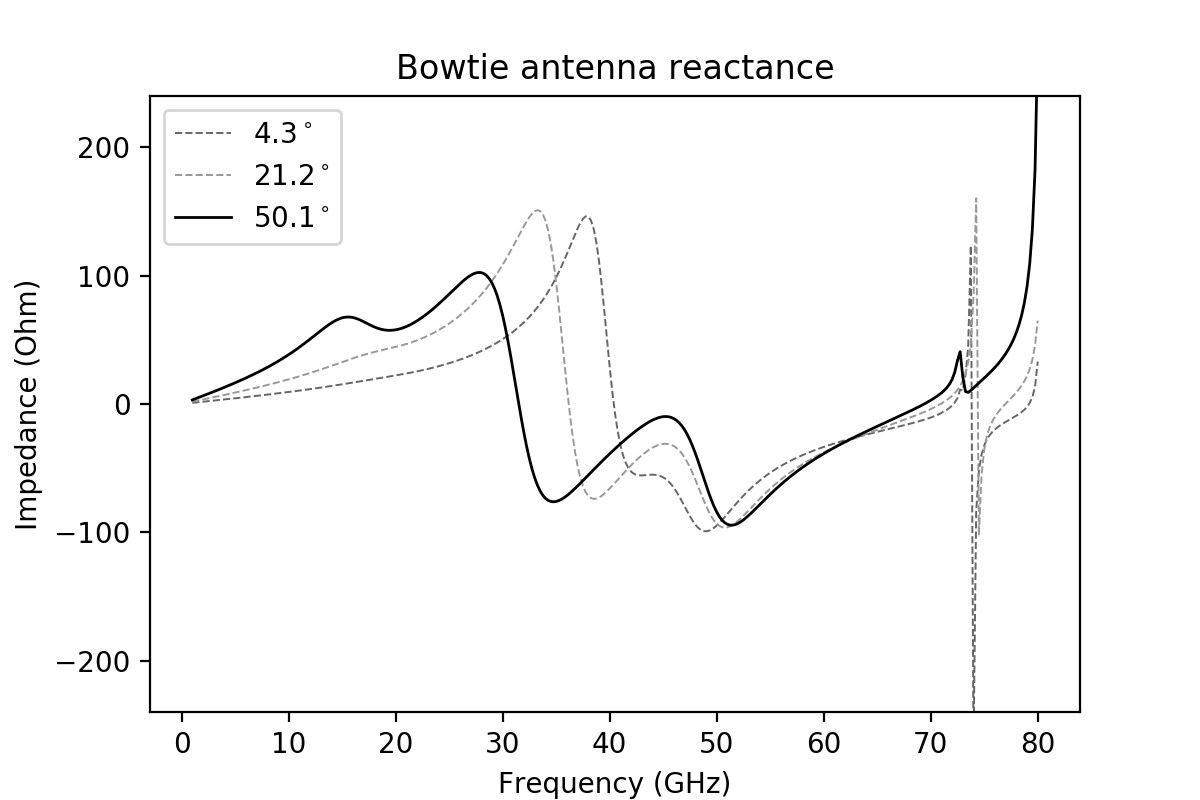}
\caption{The simulated resistance and reactance of the bowtie antenna, calculated using Ansys HFSS. The simulation shows that the antenna reactance diminishes as the flare angle increases and the first resonance widens. The solid line, labeled with a flare angle of $50^\circ$, is the nominal configuration. We drive this antenna at $100\,\Omega$ using a quasi-CPW line in our design. } 
\label{fig:antenna_impedance}
\end{figure*}
A single pixel consists of two interleaved phased antenna arrays. 
To prevent grating lobes, a phased antenna array must be spaced to Nyquist sample the focal plane \citep{kuo}. 
The antenna elements are arranged in a square lattice rotated by 45 degrees, which sets the Nyquist condition as the following inequality, 
\begin{equation} 
s \leq \frac{\lambda_{0, \min} }{\sqrt{\epsilon_r}} \left( 1- \frac{1}{N} \right) 
\label{eq:nyquist}
\end{equation}
 In this equation, $s$ is the antenna spacing and $\lambda_{0, \min}$ represents the minimum operating wavelength of the band. 
 The term in the parenthesis suppresses the array factor's end-fire beam response where $N$ is the number of antenna elements along a single axis of the square lattice. 

 This relationship imposes an upper limit on the size of the antenna element, which must not exceed a wavelength. This constraint can be challenging because wideband antennas are typically larger structures spanning multiple wavelengths \citep{chu1948}. 
Another common strategy to achieve large bandwidths is designing self-complementary antennas, which have the property of frequency-independent input impedance \citep{mushiake_1992}. This concept is the driving principle behind sinuous antennas, as detailed in \citep{roger_2008, roger_2010, roger2013}. However, this approach isn't suitable for this architecture because of space requirements for including orthogonal antenna elements. 
Consequently, implementing this phased antenna array scheme necessitates designing an antenna with a broad first resonant peak, which is then operated in that mode.  
This is the fundamental principle behind this design. 
Like the slot dipole antenna, the slotted bowtie antenna is linearly polarized.
This design incorporates a wide flare angle, which broadens the first resonance and increases the available bandwidth \citep{brown1952}.  
While a bowtie antenna is sometimes described as a traveling wave antenna parameterized solely by its flare angle, this is not true for a finite antenna lacking a resistive termination. An impedance discontinuity leads to the formation of standing waves and, therefore, resonant modes.
As discussed by \citep{stutzman, balanis}, increasing the flare angle reduces the antenna's standing wave characteristics by minimizing the phase mismatch between the voltage and currents along the slot. This results in a smoother impedance response and effectively widens the first resonance.  
This phenomenon is evident in full-field FEM simulations of the slotted bowtie antenna using Ansys High-Frequency Structure Simulator (HFSS). Figure 
\ref{fig:antenna_impedance} illustrates that an increase in flare angle shifts the first resonance to lower frequencies and smoother reactance variations. 

Additionally, rounding out the ends of the slotted bowtie modestly reduces the driving point impedance at no cost of space for our feed network \citep{qu2006}.
A flare angle of $50^\circ$ degrees was determined to meet our bandwidth requirements while still adhering to space constraints. 

In this design, depicted on the left side of Figure \ref{fig:microstrip_summingtree}, a pixel consists of a square lattice rotated by 45 degrees, highlighted in red. The antenna spacing between nearest-neighbor elements is $s = 1315\,\mu$m. 
Although this spacing is $\sim 15\%$ smaller than required by the Nyquist criteron (eq \ref{eq:nyquist}), we observed improved impedance characteristics ith the smaller spacing.

However, a more convenient mathematical representation would be the array in blue: a single polarization consists of a 12x12 square grid of resonant bowtie antenna pairs.
These pairs of antennas are arranged in a Bravais lattice with vectors $\vec{a} = (2a,0)$, and $\vec{a}_2 = (0, 2a)$, where the lattice spacing is $a = 930 \mu m$. 
Individual pairs are located at $\vec{a} = (0,0)$ and $\vec{a} = (a,\pm a)$, with the sign originating from the polarization. This alternative parameterization of the lattice is useful for the beam model in a later discussion in section \ref{sec:beams}.
The lattice for the orthogonal polarization is achieved by a simple translation of $(a,0)$.  

\subsection{Microstrip Feed Network}  
\label{sec:microstrip_fed}

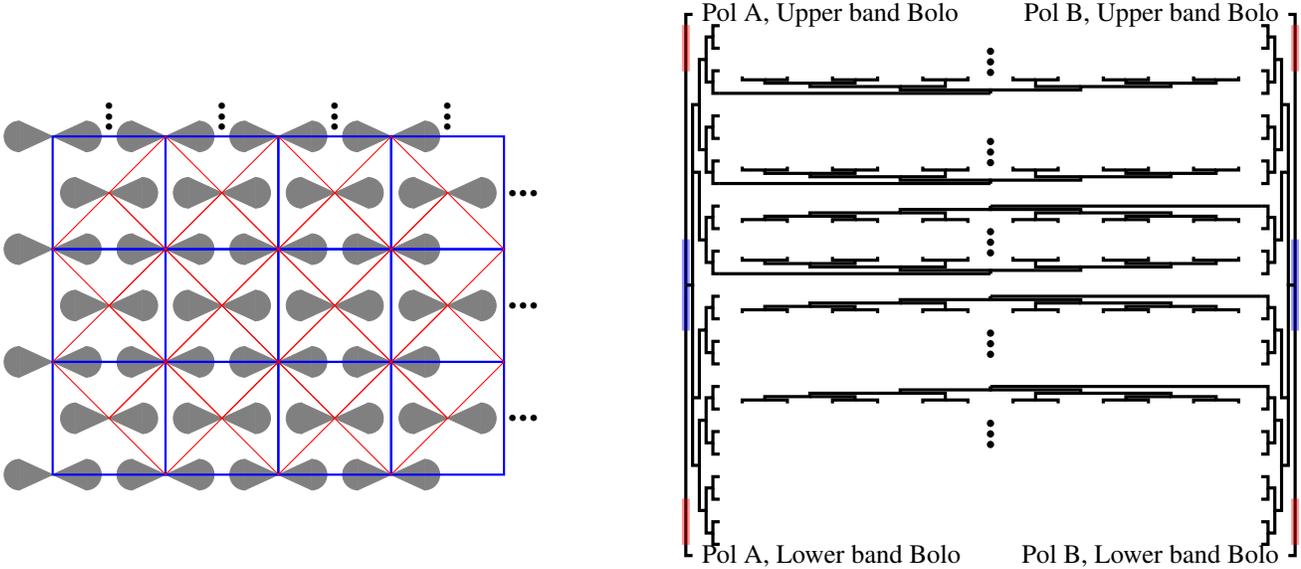
\begin{figure*}
\begin{minipage}{0.49\textwidth}


\newcommand{\bowtieAntenna}[2]{
    \begin{scope}[shift={(#1, #2)}]
	\draw [fill = gray, color = gray] (0,0) -- (-.59,.27) -- (-.59,-.27) -- cycle;
    \draw [fill = gray, color = gray](-.59,.27) arc (90:270:.27);
    \draw [fill = gray, color = gray] (0,0) -- (.59,.27) -- (.59,-.27) -- cycle;
    \draw [fill = gray, color = gray] (.59,-.27) arc (-90:90:.27);
    \end{scope}
}

\newcommand{\rotSquare}[2]{
\begin{scope}[shift={(#1,#2)}] 
 \draw[thin, red](0,0) -- (1, 1) -- (2, 0) -- (1, -1) -- cycle; 
\end{scope}
}
\begin{tikzpicture}[scale = 0.75]
    \bowtieAntenna{0}{0}
    \bowtieAntenna{2}{0}
    \bowtieAntenna{4}{0}
    \bowtieAntenna{6}{0}
    \bowtieAntenna{1}{1}
    \bowtieAntenna{3}{1}
    \bowtieAntenna{5}{1}
    \bowtieAntenna{7}{1}
    
    \bowtieAntenna{0}{2}
    \bowtieAntenna{2}{2}
    \bowtieAntenna{4}{2}
    \bowtieAntenna{6}{2}
    \bowtieAntenna{1}{3}
    \bowtieAntenna{3}{3}
    \bowtieAntenna{5}{3}
    \bowtieAntenna{7}{3}
    \bowtieAntenna{0}{4}
    \bowtieAntenna{2}{4}
    \bowtieAntenna{4}{4}
    \bowtieAntenna{6}{4}
    \bowtieAntenna{1}{5}
    \bowtieAntenna{3}{5}
    \bowtieAntenna{5}{5}
    \bowtieAntenna{7}{5}
    \bowtieAntenna{0}{6}
    \bowtieAntenna{2}{6}
    \bowtieAntenna{4}{6}
    \bowtieAntenna{6}{6}

    \draw[thick, blue](0,0)  rectangle(2,2); 
    \draw[thick, blue](2,0)  rectangle(4,2); 
    \draw[thick, blue](4,0)  rectangle(6,2); 
    \draw[thick, blue](6,0)  rectangle(8,2); 
    \draw[thick, blue](0,2)  rectangle(2,4); 
    \draw[thick, blue](2,2)  rectangle(4,4); 
    \draw[thick, blue](4,2)  rectangle(6,4); 
    \draw[thick, blue](6,2)  rectangle(8,4); 
    
    \draw[thick, blue](0,4)  rectangle(2,6); 
    \draw[thick, blue](2,4)  rectangle(4,6); 
    \draw[thick, blue](4,4)  rectangle(6,6); 
    \draw[thick, blue](6,4)  rectangle(8,6);

    \rotSquare{1}{1}
    \rotSquare{3}{1}
    \rotSquare{5}{1}
    \rotSquare{0}{2}
    \rotSquare{2}{2}
    \rotSquare{4}{2}
    \rotSquare{6}{2}
    \rotSquare{1}{3}
    \rotSquare{3}{3}
    \rotSquare{5}{3}
    
    \rotSquare{0}{4}
    \rotSquare{2}{4}
    \rotSquare{4}{4}
    \rotSquare{6}{4}
    \rotSquare{1}{5}
    \rotSquare{3}{5}
    \rotSquare{5}{5}

    \draw(8.2,1) node[rotate = 90]{\Large \bfseries $\vdots$}; 
    \draw(8.2,3) node[rotate = 90]{\Large \bfseries $\vdots$}; 
    \draw(8.2,5) node[rotate = 90]{\Large \bfseries $\vdots$}; 
    \draw(1, 6.5) node[]{\Large \bfseries $\vdots$};
    \draw(3, 6.5) node[]{\Large \bfseries $\vdots$};
    \draw(5, 6.5) node[]{\Large \bfseries $\vdots$}; 
    
    \draw(7, 6.5) node[]{\Large \bfseries $\vdots$};
\end{tikzpicture}
\end{minipage}
\begin{minipage}{0.5\textwidth}
\begin{tikzpicture}[scale = 0.3]

\draw[very thick](\A, 0) -- (\A, 12) -- (\B, 12) node[right, font = \normalsize]{Pol A, Upper band Bolo}; 
\draw[very thick](\A, 0) -- (\A, -12) -- (\B, -12)node[right, font = \normalsize]{Pol A, Lower band Bolo}; 
\draw[draw = blue, fill = blue, opacity = 0.4](-0.15, -2) rectangle ++(0.3, 4); 
\draw[draw = red, fill = red, opacity = 0.4](-0.15, -9.5) rectangle ++(0.3, -2); 
\draw[draw = red, fill = red, opacity = 0.4](-0.15, 9.5) rectangle ++(0.3, 2); 

\draw[very thick] (\A,0) -- (\B, 0) -- (\B, 5) -- (\C, 5) -- (\C, 2) -- (\D,2) -- (\D,1) -- (\E, 1) -- (\E, 0.5) -- (\F, 0.5) ; 
\draw[very thick] (\E, 1) -- (\E, 1.5) -- (\F, 1.5); 
\draw[very thick] (\D, 2) -- (\D, 3) -- (\E, 3) -- (\E, 2.5) -- (\F, 2.5); 
\draw[very thick](\E, 3) -- (\E, 3.5) -- (\F, 3.5); 
\draw[very thick](\B,5) -- (\B,8) -- (\C, 8) -- (\C, 6) -- (\D, 6) -- (\D, 5) -- (\E, 5) -- (\E, 4.5) -- (\F, 4.5);
\draw[very thick](\E, 5) -- (\E, 5.5) -- (\F, 5.5); 
\draw[very thick](\D,6) -- (\D, 7) -- (\E, 7) -- (\E, 6.5) -- (\F, 6.5); 
\draw[very thick](\E, 7) -- (\E, 7.5) -- (\F, 7.5); 
\draw[very thick](\C, 8) -- (\C, 10) -- (\D, 10) -- (\D, 9) -- (\E, 9) -- (\E, 8.5) -- (\F, 8.5); 
\draw[very thick](\E, 9) -- (\E, 9.5) -- (\F, 9.5);
\draw[very thick](\D, 10) -- (\D, 11) -- (\E, 11) -- (\E, 10.5) -- (\F, 10.5); 
\draw[very thick](\E, 11) -- (\E, 11.5) -- (\F, 11.5); 

\draw[very thick] (\A,0) -- (\B, 0) -- (\B, -5) -- (\C, -5) -- (\C, -2) -- (\D,-2) -- (\D,-1) -- (\E, -1) -- (\E, -0.5) -- (\F, -0.5) ; 
\draw[very thick] (\E, -1) -- (\E, -1.5) -- (\F, -1.5); 
\draw[very thick] (\D, -2) -- (\D, -3) -- (\E, -3) -- (\E, -2.5) -- (\F, -2.5); 
\draw[very thick](\E, -3) -- (\E, -3.5) -- (\F, -3.5); 
\draw[very thick](\B,-5) -- (\B,-8) -- (\C, -8) -- (\C, -6) -- (\D, -6) -- (\D,- 5) -- (\E, -5) -- (\E, -4.5) -- (\F, -4.5);
\draw[very thick](\E, -5) -- (\E, -5.5) -- (\F, -5.5); 
\draw[very thick](\D,-6) -- (\D, -7) -- (\E, -7) -- (\E, -6.5) -- (\F, -6.5); 
\draw[very thick](\E, -7) -- (\E, -7.5) -- (\F, -7.5); 
\draw[very thick](\C, -8) -- (\C, -10) -- (\D, -10) -- (\D, -9) -- (\E, -9) -- (\E, -8.5) -- (\F, -8.5); 
\draw[very thick](\E, -9) -- (\E, -9.5) -- (\F, -9.5);
\draw[very thick](\D, -10) -- (\D, -11) -- (\E, -11) -- (\E, -10.5) -- (\F, -10.5); 
\draw[very thick](\E, -11) -- (\E, -11.5) -- (\F, -11.5); 

\begin{scope}[xshift =27cm, xscale = -1]
\draw[very thick](\A, 0) -- (\A, 12) -- (\B, 12) node[left,font = \normalsize]{Pol B, Upper band Bolo}; 
\draw[very thick](\A, 0) -- (\A, -12) -- (\B, -12)node[left,font = \normalsize]{Pol B, Lower band Bolo};
\draw[draw = blue, fill = blue, opacity = 0.4](-0.15, -2) rectangle ++(0.3, 4); 
\draw[draw = red, fill = red, opacity = 0.4](-0.15, -9.5) rectangle ++(0.3, -2); 
\draw[draw = red, fill = red, opacity = 0.4](-0.15, 9.5) rectangle ++(0.3, 2);  
\draw[very thick] (\A,0) -- (\B, 0) -- (\B, 5) -- (\C, 5) -- (\C, 2) -- (\D,2) -- (\D,1) -- (\E, 1) -- (\E, 0.5) -- (\F, 0.5) ; 
\draw[very thick] (\E, 1) -- (\E, 1.5) -- (\F, 1.5); 
\draw[very thick] (\D, 2) -- (\D, 3) -- (\E, 3) -- (\E, 2.5) -- (\F, 2.5); 
\draw[very thick](\E, 3) -- (\E, 3.5) -- (\F, 3.5); 
\draw[very thick](\B,5) -- (\B,8) -- (\C, 8) -- (\C, 6) -- (\D, 6) -- (\D, 5) -- (\E, 5) -- (\E, 4.5) -- (\F, 4.5);
\draw[very thick](\E, 5) -- (\E, 5.5) -- (\F, 5.5); 
\draw[very thick](\D,6) -- (\D, 7) -- (\E, 7) -- (\E, 6.5) -- (\F, 6.5); 
\draw[very thick](\E, 7) -- (\E, 7.5) -- (\F, 7.5); 
\draw[very thick](\C, 8) -- (\C, 10) -- (\D, 10) -- (\D, 9) -- (\E, 9) -- (\E, 8.5) -- (\F, 8.5); 
\draw[very thick](\E, 9) -- (\E, 9.5) -- (\F, 9.5);
\draw[very thick](\D, 10) -- (\D, 11) -- (\E, 11) -- (\E, 10.5) -- (\F, 10.5); 
\draw[very thick](\E, 11) -- (\E, 11.5) -- (\F, 11.5); 

\draw[very thick] (\A,0) -- (\B, 0) -- (\B, -5) -- (\C, -5) -- (\C, -2) -- (\D,-2) -- (\D,-1) -- (\E, -1) -- (\E, -0.5) -- (\F, -0.5) ; 
\draw[very thick] (\E, -1) -- (\E, -1.5) -- (\F, -1.5); 
\draw[very thick] (\D, -2) -- (\D, -3) -- (\E, -3) -- (\E, -2.5) -- (\F, -2.5); 
\draw[very thick](\E, -3) -- (\E, -3.5) -- (\F, -3.5); 
\draw[very thick](\B,-5) -- (\B,-8) -- (\C, -8) -- (\C, -6) -- (\D, -6) -- (\D,- 5) -- (\E, -5) -- (\E, -4.5) -- (\F, -4.5);
\draw[very thick](\E, -5) -- (\E, -5.5) -- (\F, -5.5); 
\draw[very thick](\D,-6) -- (\D, -7) -- (\E, -7) -- (\E, -6.5) -- (\F, -6.5); 
\draw[very thick](\E, -7) -- (\E, -7.5) -- (\F, -7.5); 
\draw[very thick](\C, -8) -- (\C, -10) -- (\D, -10) -- (\D, -9) -- (\E, -9) -- (\E, -8.5) -- (\F, -8.5); 
\draw[very thick](\E, -9) -- (\E, -9.5) -- (\F, -9.5);
\draw[very thick](\D, -10) -- (\D, -11) -- (\E, -11) -- (\E, -10.5) -- (\F, -10.5); 
\draw[very thick](\E, -11) -- (\E, -11.5) -- (\F, -11.5); 
\end{scope}
\begin{scope}[xshift = 13.5cm, yshift=0.5cm, yscale = 0.5, xscale = 2]
\draw[very thick](0,0) -- (-6, 0); 
\draw[very thick](0,\Z1) -- (0, \Z2) -- (-2, \Z2) -- (-2, \Z3) -- (-4, \Z3) -- (-4, \Z4) -- (-5, \Z4) -- (-5, \Z5) -- (-5.5, \Z5) -- (-5.5, \Z6);
\draw[very thick](-5, \Z5) -- (-4.5, \Z5) -- (-4.5, \Z6); 
\draw[very thick](-4, \Z4) -- (-3, \Z4) -- (-3, \Z5) -- (-3.5, \Z5) -- (-3.5,\Z6); 
\draw[very thick](-3, \Z5) -- (-2.5, \Z5) -- (-2.5, \Z6); 
\draw[very thick](-2, \Z3) -- (-1, \Z3) -- (-1, \Z5) -- (-1.5, \Z5) -- (-1.5, \Z6); 
\draw[very thick](-1, \Z5) -- (-0.5, \Z5) -- (-0.5, \Z6); 
\draw[very thick](2, \Z3) -- (1, \Z3) -- (1, \Z5) -- (1.5, \Z5) -- (1.5, \Z6); 
\draw[very thick](1, \Z5) -- (0.5, \Z5) -- (0.5, \Z6); 
\draw[very thick](0,\Z1) -- (0, \Z2) -- (2, \Z2) -- (2, \Z3) -- (4, \Z3) -- (4, \Z4) -- (5, \Z4) -- (5, \Z5) -- (5.5, \Z5) -- (5.5, \Z6);
\draw[very thick](5, \Z5) -- (4.5, \Z5) -- (4.5, \Z6); 
\draw[very thick](4, \Z4) -- (3, \Z4) -- (3, \Z5) -- (3.5, \Z5) -- (3.5,\Z6); 
\draw[very thick](3, \Z5) -- (2.5, \Z5) -- (2.5, \Z6); 
\draw(0, \Z6 + 2 )node[font = \Huge]{$\vdots$}; 
\end{scope}

\begin{scope}[xshift = 13.5cm, yshift=4.5cm, yscale = 0.5, xscale = 2]
\draw[very thick](0,0) -- (-6, 0); 
\draw[very thick](0,\Z1) -- (0, \Z2) -- (-2, \Z2) -- (-2, \Z3) -- (-4, \Z3) -- (-4, \Z4) -- (-5, \Z4) -- (-5, \Z5) -- (-5.5, \Z5) -- (-5.5, \Z6);
\draw[very thick](-5, \Z5) -- (-4.5, \Z5) -- (-4.5, \Z6); 
\draw[very thick](-4, \Z4) -- (-3, \Z4) -- (-3, \Z5) -- (-3.5, \Z5) -- (-3.5,\Z6); 
\draw[very thick](-3, \Z5) -- (-2.5, \Z5) -- (-2.5, \Z6); 
\draw[very thick](-2, \Z3) -- (-1, \Z3) -- (-1, \Z5) -- (-1.5, \Z5) -- (-1.5, \Z6); 
\draw[very thick](-1, \Z5) -- (-0.5, \Z5) -- (-0.5, \Z6); 
\draw[very thick](2, \Z3) -- (1, \Z3) -- (1, \Z5) -- (1.5, \Z5) -- (1.5, \Z6); 
\draw[very thick](1, \Z5) -- (0.5, \Z5) -- (0.5, \Z6); 
\draw[very thick](0,\Z1) -- (0, \Z2) -- (2, \Z2) -- (2, \Z3) -- (4, \Z3) -- (4, \Z4) -- (5, \Z4) -- (5, \Z5) -- (5.5, \Z5) -- (5.5, \Z6);
\draw[very thick](5, \Z5) -- (4.5, \Z5) -- (4.5, \Z6); 
\draw[very thick](4, \Z4) -- (3, \Z4) -- (3, \Z5) -- (3.5, \Z5) -- (3.5,\Z6); 
\draw[very thick](3, \Z5) -- (2.5, \Z5) -- (2.5, \Z6); 
\draw(0, \Z6 + 2 )node[font = \Huge]{$\vdots$}; 
\end{scope}

\begin{scope}[xshift = 13.5cm, yshift=8.5cm, yscale = 0.5, xscale = 2]
\draw[very thick](0,0) -- (-6, 0); 
\draw[very thick](0,\Z1) -- (0, \Z2) -- (-2, \Z2) -- (-2, \Z3) -- (-4, \Z3) -- (-4, \Z4) -- (-5, \Z4) -- (-5, \Z5) -- (-5.5, \Z5) -- (-5.5, \Z6);
\draw[very thick](-5, \Z5) -- (-4.5, \Z5) -- (-4.5, \Z6); 
\draw[very thick](-4, \Z4) -- (-3, \Z4) -- (-3, \Z5) -- (-3.5, \Z5) -- (-3.5,\Z6); 
\draw[very thick](-3, \Z5) -- (-2.5, \Z5) -- (-2.5, \Z6); 
\draw[very thick](-2, \Z3) -- (-1, \Z3) -- (-1, \Z5) -- (-1.5, \Z5) -- (-1.5, \Z6); 
\draw[very thick](-1, \Z5) -- (-0.5, \Z5) -- (-0.5, \Z6); 
\draw[very thick](2, \Z3) -- (1, \Z3) -- (1, \Z5) -- (1.5, \Z5) -- (1.5, \Z6); 
\draw[very thick](1, \Z5) -- (0.5, \Z5) -- (0.5, \Z6); 
\draw[very thick](0,\Z1) -- (0, \Z2) -- (2, \Z2) -- (2, \Z3) -- (4, \Z3) -- (4, \Z4) -- (5, \Z4) -- (5, \Z5) -- (5.5, \Z5) -- (5.5, \Z6);
\draw[very thick](5, \Z5) -- (4.5, \Z5) -- (4.5, \Z6); 
\draw[very thick](4, \Z4) -- (3, \Z4) -- (3, \Z5) -- (3.5, \Z5) -- (3.5,\Z6); 
\draw[very thick](3, \Z5) -- (2.5, \Z5) -- (2.5, \Z6); 
\draw(0, \Z6 + 2 )node[font = \Huge]{$\vdots$}; 
\end{scope}


\begin{scope}[xshift = 13.5cm, yshift=-0.5cm, yscale = 0.5, xscale = 2]
\draw[very thick](0,0) -- (6, 0); 
\draw[very thick](0,\Y1) -- (0, \Y2) -- (-2, \Y2) -- (-2, \Y3) -- (-4, \Y3) -- (-4, \Y4) -- (-5, \Y4) -- (-5, \Y5) -- (-5.5, \Y5) -- (-5.5, \Y6);
\draw[very thick](-5, \Y5) -- (-4.5, \Y5) -- (-4.5, \Y6); 
\draw[very thick](-4, \Y4) -- (-3, \Y4) -- (-3, \Y5) -- (-3.5, \Y5) -- (-3.5,\Y6); 
\draw[very thick](-3, \Y5) -- (-2.5, \Y5) -- (-2.5, \Y6); 
\draw[very thick](-2, \Y3) -- (-1, \Y3) -- (-1, \Y5) -- (-1.5, \Y5) -- (-1.5, \Y6); 
\draw[very thick](-1, \Y5) -- (-0.5, \Y5) -- (-0.5, \Y6); 
\draw[very thick](2, \Y3) -- (1, \Y3) -- (1, \Y5) -- (1.5, \Y5) -- (1.5, \Y6); 
\draw[very thick](1, \Y5) -- (0.5, \Y5) -- (0.5, \Y6); 
\draw[very thick](0,\Y1) -- (0, \Y2) -- (2, \Y2) -- (2, \Y3) -- (4, \Y3) -- (4, \Y4) -- (5, \Y4) -- (5, \Y5) -- (5.5, \Y5) -- (5.5, \Y6);
\draw[very thick](5, \Y5) -- (4.5, \Y5) -- (4.5, \Y6); 
\draw[very thick](4, \Y4) -- (3, \Y4) -- (3, \Y5) -- (3.5, \Y5) -- (3.5,\Y6); 
\draw[very thick](3, \Y5) -- (2.5, \Y5) -- (2.5, \Y6); 
\draw(0, \Y6 - 2 )node[font = \Huge]{$\vdots$}; 
\end{scope}  

\begin{scope}[xshift = 13.5cm, yshift=3.5cm, yscale = 0.5, xscale = 2]
\draw[very thick](0,0) -- (6, 0); 
\draw[very thick](0,\Y1) -- (0, \Y2) -- (-2, \Y2) -- (-2, \Y3) -- (-4, \Y3) -- (-4, \Y4) -- (-5, \Y4) -- (-5, \Y5) -- (-5.5, \Y5) -- (-5.5, \Y6);
\draw[very thick](-5, \Y5) -- (-4.5, \Y5) -- (-4.5, \Y6); 
\draw[very thick](-4, \Y4) -- (-3, \Y4) -- (-3, \Y5) -- (-3.5, \Y5) -- (-3.5,\Y6); 
\draw[very thick](-3, \Y5) -- (-2.5, \Y5) -- (-2.5, \Y6); 
\draw[very thick](-2, \Y3) -- (-1, \Y3) -- (-1, \Y5) -- (-1.5, \Y5) -- (-1.5, \Y6); 
\draw[very thick](-1, \Y5) -- (-0.5, \Y5) -- (-0.5, \Y6); 
\draw[very thick](2, \Y3) -- (1, \Y3) -- (1, \Y5) -- (1.5, \Y5) -- (1.5, \Y6); 
\draw[very thick](1, \Y5) -- (0.5, \Y5) -- (0.5, \Y6); 
\draw[very thick](0,\Y1) -- (0, \Y2) -- (2, \Y2) -- (2, \Y3) -- (4, \Y3) -- (4, \Y4) -- (5, \Y4) -- (5, \Y5) -- (5.5, \Y5) -- (5.5, \Y6);
\draw[very thick](5, \Y5) -- (4.5, \Y5) -- (4.5, \Y6); 
\draw[very thick](4, \Y4) -- (3, \Y4) -- (3, \Y5) -- (3.5, \Y5) -- (3.5,\Y6); 
\draw[very thick](3, \Y5) -- (2.5, \Y5) -- (2.5, \Y6); 
\end{scope}  
\begin{scope}[xshift = 13.5cm, yshift=-4.5cm, yscale = 0.5, xscale = 2]
\draw[very thick](0,0) -- (6, 0); 
\draw[very thick](0,\Y1) -- (0, \Y2) -- (-2, \Y2) -- (-2, \Y3) -- (-4, \Y3) -- (-4, \Y4) -- (-5, \Y4) -- (-5, \Y5) -- (-5.5, \Y5) -- (-5.5, \Y6);
\draw[very thick](-5, \Y5) -- (-4.5, \Y5) -- (-4.5, \Y6); 
\draw[very thick](-4, \Y4) -- (-3, \Y4) -- (-3, \Y5) -- (-3.5, \Y5) -- (-3.5,\Y6); 
\draw[very thick](-3, \Y5) -- (-2.5, \Y5) -- (-2.5, \Y6); 
\draw[very thick](-2, \Y3) -- (-1, \Y3) -- (-1, \Y5) -- (-1.5, \Y5) -- (-1.5, \Y6); 
\draw[very thick](-1, \Y5) -- (-0.5, \Y5) -- (-0.5, \Y6); 
\draw[very thick](2, \Y3) -- (1, \Y3) -- (1, \Y5) -- (1.5, \Y5) -- (1.5, \Y6); 
\draw[very thick](1, \Y5) -- (0.5, \Y5) -- (0.5, \Y6); 
\draw[very thick](0,\Y1) -- (0, \Y2) -- (2, \Y2) -- (2, \Y3) -- (4, \Y3) -- (4, \Y4) -- (5, \Y4) -- (5, \Y5) -- (5.5, \Y5) -- (5.5, \Y6);
\draw[very thick](5, \Y5) -- (4.5, \Y5) -- (4.5, \Y6); 
\draw[very thick](4, \Y4) -- (3, \Y4) -- (3, \Y5) -- (3.5, \Y5) -- (3.5,\Y6); 
\draw[very thick](3, \Y5) -- (2.5, \Y5) -- (2.5, \Y6); 
\draw(0, \Y6 - 2 )node[font = \Huge]{$\vdots$}; 
\end{scope}  
\end{tikzpicture} 
\end{minipage}
\caption{ (Left) An abbreviated schematic of a single-polarization antenna array, depicted in two colors to illustrate two possible descriptions. The red array represents this design as a square array rotated by 45 degrees, establishing the Nyquist criterion for antenna spacing. Alternatively, the blue array describes our configuration as a product of a square array with two elements per square, offering a convenient analytical description of the total beam. (Right) An abbreviated schematic of a single pixel's microstrip summing tree network. Antenna sub-element are located at the ends of all horizontal brackets. Power is coherently summed by the horizontal network, and then summed by the vertical network. Different polarizations are directed left and right. The signal is partitioned by a diplexer, as depicted in blue, and filtered by a bandpass filter, as depicted in red, and terminated on a bolometer island. }
\label{fig:microstrip_summingtree}
\end{figure*}

We collect power by combining signals from 12$\times$12 bowtie pairs in the array with a microstrip feed network.  All antennas are coherently fed with uniform power division. 
A cartoon diagram of a simplified microstrip network is shown by the right side of Figure \ref{fig:microstrip_summingtree}. 
We achieve coherent summation by summing waves coherently by row and then as a column using microstrip T-junctions.
 The power is subsequently split by a diplexer and filtered before terminating onto a bolometer island.  

 We refer to uniform and in-phase summation as \emph{top-hat} illumination. 
We chose top-hat illumination for its simplicity. However, it's important to note that tapered illumination could improve beam quality \citep{antenna_coupled_tes_new}, as top-hat illumination tends to result in larger side lobes.

A relatively high impedance line of around $\sim 100\,\Omega$ is required to drive the bowtie antenna efficiently. 
This impedance is unachievable with microstrip lines with a thin ILD ($0.3\,\mu m$ of SiO$_2$). 
Instead, we can create a high impedance line by removing the ground plane directly underneath the conductor, so that the field lines mimic those of a CPW \citep{quasicpw}.
We found that a substantial gap, $20\,\mu$m side-to-side, allowed us to achieve our desired impedance. Then, adjacent quasi-CPW lines join at a T-junction and transition, with minimal reflection loss, to a microstrip line for the remainder of the summing tree. 

Furthermore, ground bridges are strategically placed to short the two outer ground conductors.  
These bridges eliminate potential slotline modes that may radiate. Their spacing has determined to ensure that no slotline modes are excited within the operating bandwidth of this detector array. 

Crosstalk can be a concern with phased array antennas. Coupling between neighboring transmission lines can introduce phase errors across the pixel, leading to beam steering \citep{antenna_coupled_tes_bolo}. The larger footprint of the bowtie antenna restricts the spacing available for routing the summing tree. 
We mitigate crosstalk by two design principles. Firstly, we opt for a thin ILD to enhance the confinement of the field lines of the microstrip mode. Secondly, we route the microstrip lines to fan in and out wherever space allows.
Coupling between microstrip lines relies on excitations of even and odd modes \citep{pozar}. 
By flaring the microstrip lines, no significant sections of the summing tree remain in close contact, and the two modes continually change impedances along the lines. This prevents coupling even when lines must compress within a few linewidths of distance. 
The spacing of the antennas necessitate lines as close as 5\,$\mu$m. However, these modes are confined within a thin ILD (0.3\,$\mu$m) and remain in close proximity for no more than 500\,$\mu$m, about 10\% of a wavelength, resulting in minimal coupling. HFSS simulations show crosstalk levels of $-50\,$dB between neighboring lines at our tightest junctions.
We expect cross-talk to be a minor contributor to beam effects, with frame-edge coupling assuming the predominant role.
 
 \subsection{On-Chip Filtering} 
 \label{sec:filtering}
\begin{figure*}
\begin{minipage}{0.4\textwidth}
\includegraphics[width = \textwidth, height = 2cm]{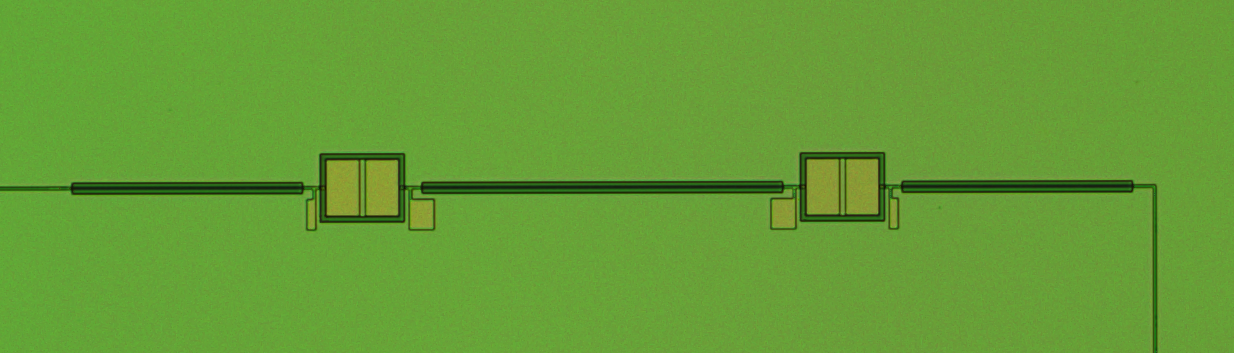}
\end{minipage}\begin{minipage}{0.6\textwidth}
 \includegraphics[width = \textwidth, height = 2.2cm]{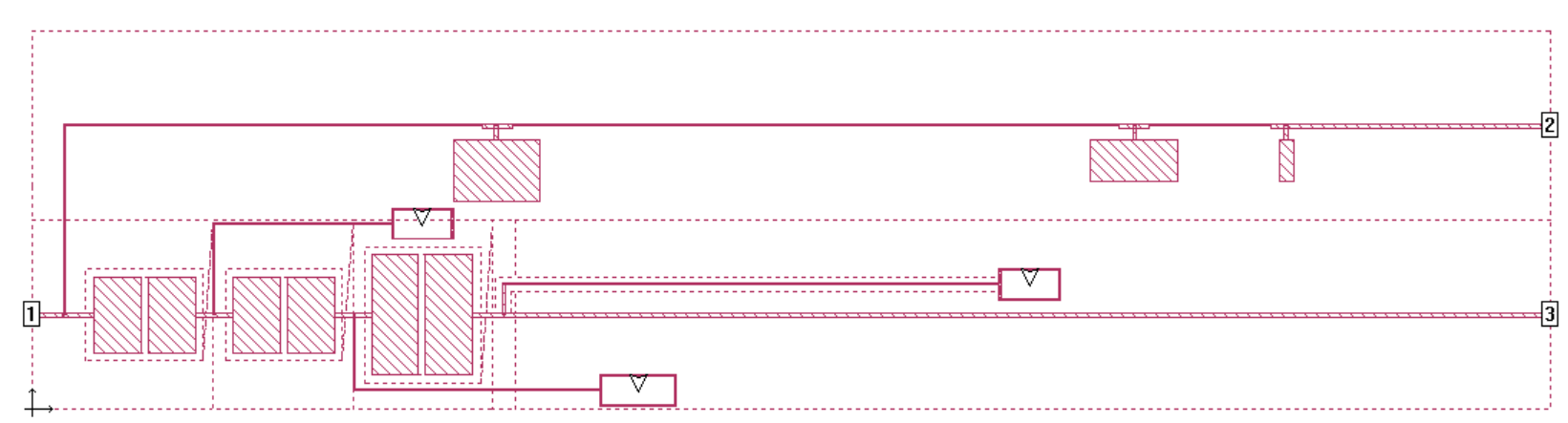} 
 \end{minipage}

\vspace{-0.5ex} 
\begin{minipage}{0.4\textwidth}
\begin{circuitikz}[scale = 0.6, transform shape, /tikz/circuitikz/bipoles/length=0.9cm]
\draw(0,0) to[L, l = $L_1$, o-*](2,0); 
\draw(2,0) to[C, l_ = $C_1$, -](2,-1.5) node[ground]{};
\draw(2,0) to[C, l = $C_2$, -](3.5,0) to[C, l = $C_3$](3.5,-1.5) node[ground]{}; 
\draw(3.5,0) to[L, l = $L_2$, *-*](6.5,0); 
\draw(6.5,0) to[C, l_= $C_3$](6.5, -1.5) node[ground]{};
\draw(6.5,0) to[C, l = $C_2$](8,0); 
\draw(8,0) to[C, l = $C_1$](8,-1.5) node[ground]{}; 
\draw(8,0) to[L, l = $L_1$, *-o](10,0); 
\end{circuitikz} 
\end{minipage}\begin{minipage}{0.6\textwidth}
 \begin{circuitikz}[scale = 0.7, transform shape, /tikz/circuitikz/bipoles/length = 0.9cm] 
 
 \draw(0,0.5) to [short, o-](0,2) to[short](0.5,2); 
 \draw(0.5,2) to[C, l = $\frac{1}{L_1}$](2,2) to[L, l = $\frac{1}{C_1}$](2, 1) node[ground]{}; 
 \draw(2,2) to[C, l = $\frac{1}{L_2}$](3.5, 2) to[L, l = $\frac{1}{C_2}$](3.5, 1) node[ground]{}; 
 \draw(3.5, 2) to[C, l = $\frac{1}{L_3}$] (5,2) to[L, l =$\frac{1}{C_3}$](5, 1) node[ground]{}; 
 \draw(5, 2) to[short, -o] (6, 2); 
 \draw(6, 2.2) node{High pass};

 \draw(0,2) to[short](-0.5,2); 
 \draw(-0.5,2) to[L, l_ = $L_1$](-2,2) to[C, l_ = $C_1$](-2, 1) node[ground]{}; 
 \draw(-2,2) to[L, l_ = $L_2$](-3.5, 2) to[C, l_ = $C_2$](-3.5, 1) node[ground]{}; 
 \draw(-3.5, 2) to[L, l_ = $L_3$] (-5,2) to[C, l_ =$C_3$](-5, 1) node[ground]{}; 
 \draw(-5, 2) to[short, -o] (-6, 2); 
 \draw(-6, 2.2) node{Low pass}; 

 \end{circuitikz} 
 \end{minipage}

 \caption{On chip filters that define the bandpass of the detectors. On the left is the bandpass filter that defines the extremities of the bands. The bandpass filter is equivalent to a three-pole LC series resonator with impedance invertors. On the right is the diplexer composed of a Butterworth high-pass and low-pass filter. The image above is a Sonnet schematic of the diplexer, where port 1 connects to the antenna network, port 2 is the low-pass band, and port 3 is the high-pass band. Below the image is a lumped element equivalent network for this diplexer. The design table is in Table \ref{table:design}, where the inductor values must be scaled by $Z_0/\omega_0$ and capacitor values scaled by $(\omega_0 Z_0)^{-1}$.  } 
 \label{fig_filters} 
 \end{figure*}
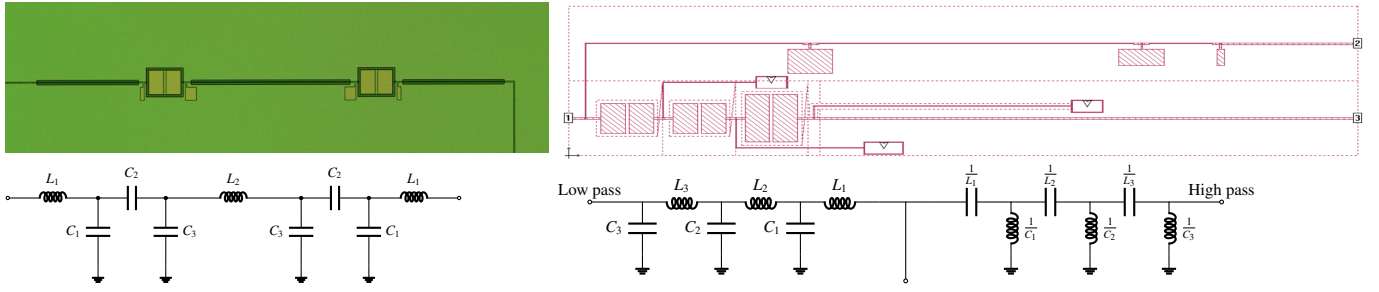
 
 \begin{table}[h] 
 \centering  
 \caption{Circuit Design Table for On-Chip Filters. The corresponding circuit elements are in Figure \ref{fig_filters}. To scale from the design table to physical values: inductor values are scaled by $\omega_0^{-1} Z_0$, and capacitor values are scaled by $(\omega_0 Z_0)^{-1}$ where $\omega_0 = 2\pi f_0$ is the desired -3dB transition between the two bands and $Z_0$ is the port impedance. For this particular design, $f_0 = 35GHz$ and $Z_0 = 25\Omega$. Simulations are then performed to convert the electrical inductance and capacitance to a lithographic element, the details of which are outlined in Appendix \ref{appendix:sonnet}. } 
 \begin{tabular}{rcccccc} \toprule  
 & $L_1$ & $C_1$ & $L_2$ & $C_2$ & $L_3$ & $C_3$ \\ \midrule 
 Bandpass & 1.85 & 0.25  & 2.70 & 0.41 & -  & 0.51 \\
 Diplexer & 1.55 & 1.76 & 1.55  & 1.20 & 0.758 &0.259  \\ \bottomrule 
 \end{tabular} 
 \label{table:design}
 \end{table}
 Each detector band is synthesized from a combination of the diplexer and the wideband bandpass filter. The bandpass filter defines the upper band edge for the high-frequency band and the lower band edge for the low-frequency band. The diplexer splits the power contiguously between the two bands. 
  
The diplexer is composed of a sixth-order high-pass and a low-pass Butterworth filter.
In this topology, the specific values for all the components have been carefully selected to achieve equal and opposite reactance for the two LC ladder circuits. Therefore, the reactance is matched across all frequencies and identically cancels out at the antenna input, ensuring a continuous power split without any reflections.  The design is described in detail in \cite{matthaei}. 
We chose this design for its simplicity and robustness to fabrication non-uniformity.

This circuit was realized by finding lithographic approximations to lumped element circuit elements through extensive simulations using Sonnet$^{\tiny \textregistered}$. Lumped inductors are synthesized by sub-wavelength high-impedance transmission lines, while lumped capacitors are synthesized by parallel plate capacitors. 
In practice, lithographic elements only approximate lumped elements at microwave frequencies. Frequency dispersion reduces their effectiveness. We selected a three-pole Butterworth filter to strike a balance between achievable lithographic elements and filter sharpness. 
In particular, series capacitors become more dispersive with larger capacitance values, nullifying the benefits of higher-order filters. 
   
The bandpass filter design concept was adapted from the previous BICEP style of filters \citep{antenna_coupled_tes_bolo}. In its most basic description, the bandpass filter architecture consists of three series LC tanks joined by impedance inverters \citep{matthaei}. 
The Kuroda identities convert the impedance K-invertor into a physically realizable capacitor network \citep{pozar}. 
Additionally, we found it advantageous to perform a $Y$ to $\pi$ transformation of the capacitor network at these microwave frequencies to reduce the series capacitances necessary for lithography. The design table is tabulated in Table \ref{table:design}. 
The left side of Figure \ref{fts_spectrum} shows the simulated performance of these filters, and we leave the discussion of the performance in Section \ref{sec:bandpass}.

\subsection{Bolometer Design}
\label{sec:bolometer}
\begin{figure*}
\begin{minipage}{0.5\textwidth}
\includegraphics[width = \textwidth]{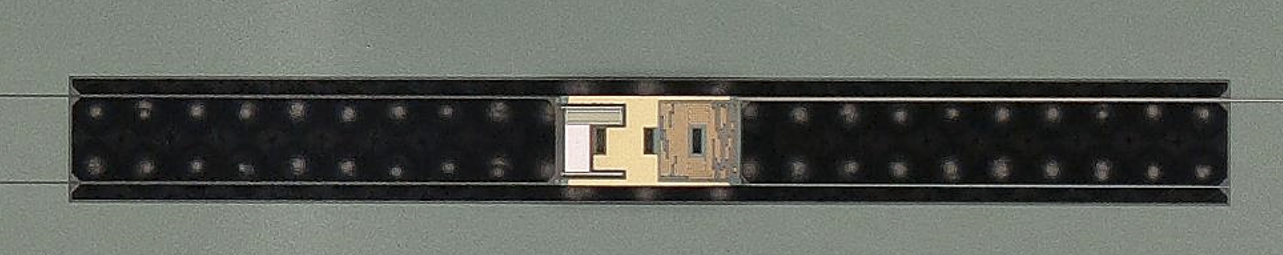}
\end{minipage}
  \begin{minipage}{0.5\textwidth}
    \centering
\begin{tabular}{rcccc}
\toprule
Bolometer Properties & $G_c$ & $T_c$ & $R_n$ & $P_{sat}$ \\ \midrule
Median & 14.8pW/K & 428mK & 101.9 m$\Omega$ & 1.58pW \\ 
Standard Deviation & 3.4pW/K & 10mK & 18.7m$\Omega$ & 0.24pW \\ \bottomrule
\end{tabular}
\end{minipage}
\caption{(Left) A microscope image of a suspended bolometer island on the detector wafer. In the photograph, on-sky power propagates from the upper right leg and is deposited on the lossy gold meander on the right side of the island. Two DC bias lines come in from the two left legs and bias the Al and Ti TES bolometers in series. (Right) A tabulation of measured bolometer properties.}
\label{table:bolo_properties}
\end{figure*}

Power captured by the antennas propagates through the microstrip network, passes through on-chip filters, and is ultimately dissipated as heat by a lossy gold termination on a suspended bolometer island. This heat is ultimately detected by a TES bolometer \citep{kent_irwin_TES}. 

TES bolometers are widely adopted in many CMB experiments because they have excellent noise properties and are a suitable choice to make large-format arrays. They can be fabricated using standard thin-film lithography techniques and, importantly, are compatible with multiplexing schemes \citep{tmux, fmux, mumux}.
Additionally, they have a strong negative electro-thermal feedback, increasing the device's linearity \citep{kent_irwin_TES}.  
A TES bolometer is operated by voltage biasing the detector between superconducting and normal states. When an incident photon is absorbed and heats the bolometer, slight temperature changes will result in significant variations in the TES resistance. 
The electrical current flowing through the device is inductively coupled to the SQUID read-out scheme, enabling a highly sensitive measurement of relative energy changes. 

The detector design has two TES in series with different superconducting transition temperatures ($T_c$). One is an aluminum TES with a $T_c = 1.2\,K$. 
This TES is designed to handle significantly higher optical loading in laboratory settings, particularly during calibration and ground-based optical characterization. 
The titanium TES has a much lower $T_c = 450\,mK$ for science observation with significantly reduced loading conditions. This TES offers superior detector stability and substantially higher sensitivity. 
%
The saturation power is a key design parameter of a bolometer and is the total power needed to bring the TES temperature to $T_{c}$: 
\begin{equation}
P_{sat} = P (T_c)= G_c T_c \frac{1-(T_{bath}/T_c)^{(n+1)}} {n+1} 
\end{equation}
where $T_{\text{bath}} \sim 280\,mK$ is the surrounding heat bath temperature, $G_{c}$ is thermal conductance of the bolometer isolation legs at $T=T_c$, and $n$ correspond to the nature of phonon transport in the legs. For our bolometer design, the isolation legs are thin, so phonons are reduced to two-dimensions and $n$ has been experimentally determined to be $\sim 2$.
$G_c$ is carefully designed so that $P_{sat}$ is above the optical loading we expect when doing science observations at the South Pole while keeping it as low as possible to minimize the noise-equivalent power (NEP). 

The noise in TES detector consists of several components, including photon noise, phonon noise, SQUID read-out noise and Johnson noise from the shunt resistor \citep{Zmuidzinas}. The latter two sources of noise can be deliberately minimized to be sub-dominant to the first two, which are intrinsic and establish the fundamental sensitivity of the bolometer. 
Thermal fluctuations across the silicon nitride legs of the bolometers contribute the following NEP, 
\begin{equation}
NEP^2_{phonon}= 4 k_B T^2_{c} G_c F(T_c,T_{bath}) 
\end{equation}
where $F(T_c,T_{bath})$ is a function of bath and TES island temperature and accounts for the non-equilibrium effects \citep{Mathern}. Typical values for our bolometers are $F \sim 0.5$. 

Because phonon NEP is proportional to the square root of thermal conductance, the value of $G_c$ has a direct impact on mapping speed. 
The bolometer island is suspended by four legs, forming the weak thermal link between the island and the thermal bath. The $G_c$, tunable by leg lengths, is optimally designed to maximize detector sensitivity, while avoiding saturation during science operation. The bolometer legs are constructed from $1\,\mu m$ thick low-stress nitride (LSN) and are $9\,\mu m$ wide for the leg that bridges the wire connecting microstrip suming tree to the resistive termination, $6\,\mu m$ wide for two legs that carry the DC bias lines, and $4\, \mu m$ thick for one remaining support leg. The length legs are 800$\,\mu m$ long. Note that these differ from the $G_c$ values used for the monochromatic 30/40\,GHz bolometer arrays as described in other publications \citep{cheng, cheng_thesis}. 
A tabulation of the bolometer properties and performance is shown in Figure \ref{table:bolo_properties}. 
While detector sensitivity can be enhanced by reducing phonon noise with a lower $T_{\text{bath}} = 100\,$mK, the incremental gains are outweighed by the increased cryogenic complexity, given that these detectors are photon-noise limited. Consequently, a lower 100\,mK design was not pursued \citep{cheng}.

\FloatBarrier 
\section{Optical Characterization}  

This focal plane was cooled down and tested in a testbed cryostat that mimics the filtering present on a BICEP-style receiver but lacks the imaging optics of the complete telescope insert. 
A series of optical filters, including absorbing plastic filters and reflective metal-mesh filters, was used to limit the radiative thermal load on the focal plane\citep{ade_filters}. 
The on-chip filtering ultimately defines the bands of the detectors for science observation. All laboratory measurements were performed using the aluminum superconducting transition designed for higher optical load. 

 To minimize radio frequency interference (RFI), archival measurements were conducted at night with WiFi routers and personal wireless devices powered off. 
 Laboratory testing has demonstrated that majority of RFI for these style receivers couples in through the readout chain rather than the optical chain \citep{Ahmed_thesis}. To reduce RF pick-up, a Faraday cage made of 1/8" wire metal mesh enclosed the readout electronics and housekeeping electronics. Additionally, aluminumized Mylar tape is used between readout modules for RF shielding.

 \subsection{Spectral Bandpass Characterization}
\label{sec:bandpass}
\begin{figure*}
%
\centering
\includegraphics[width = 0.45\textwidth]{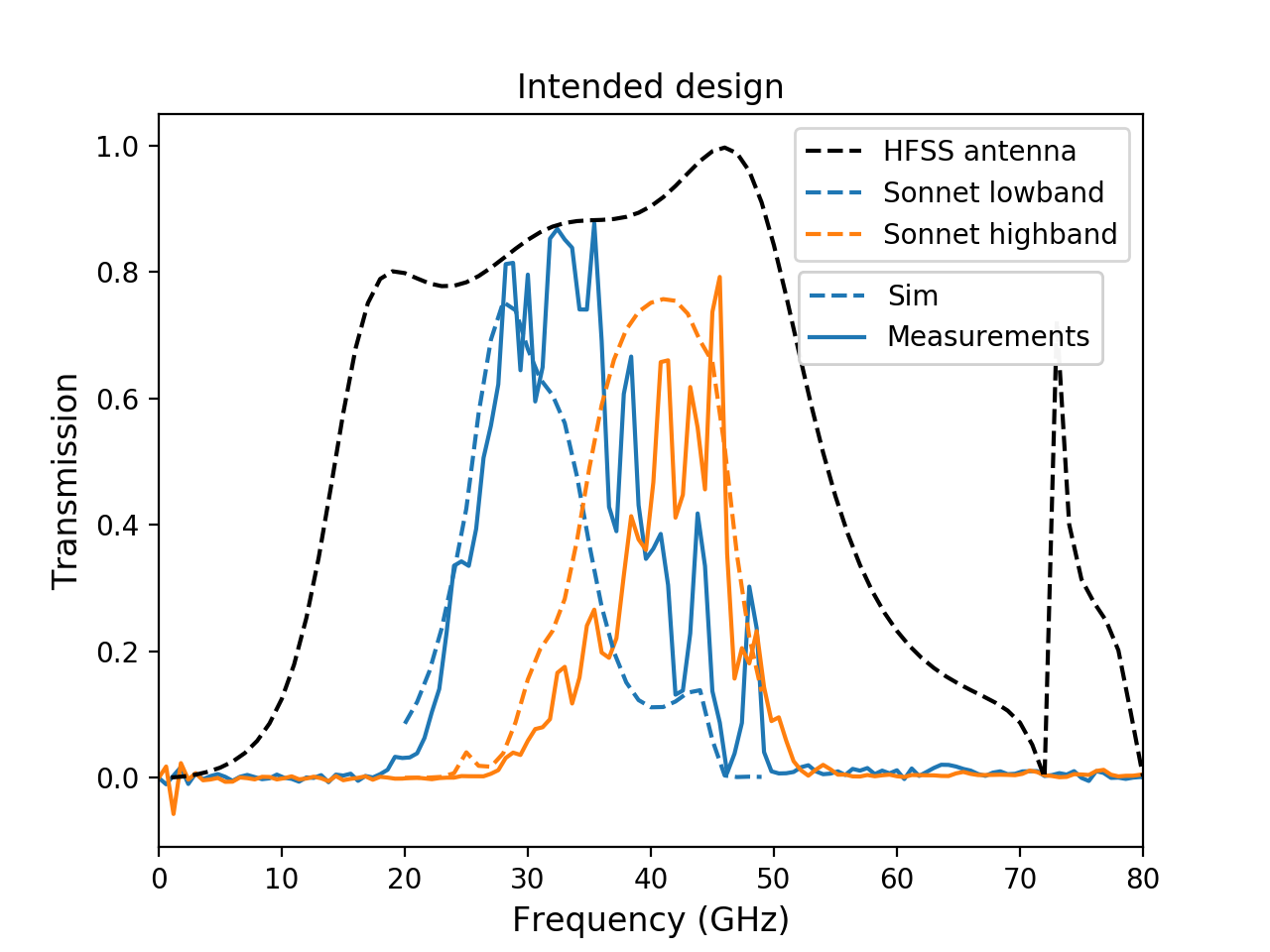} 
\includegraphics[width = 0.45\textwidth]{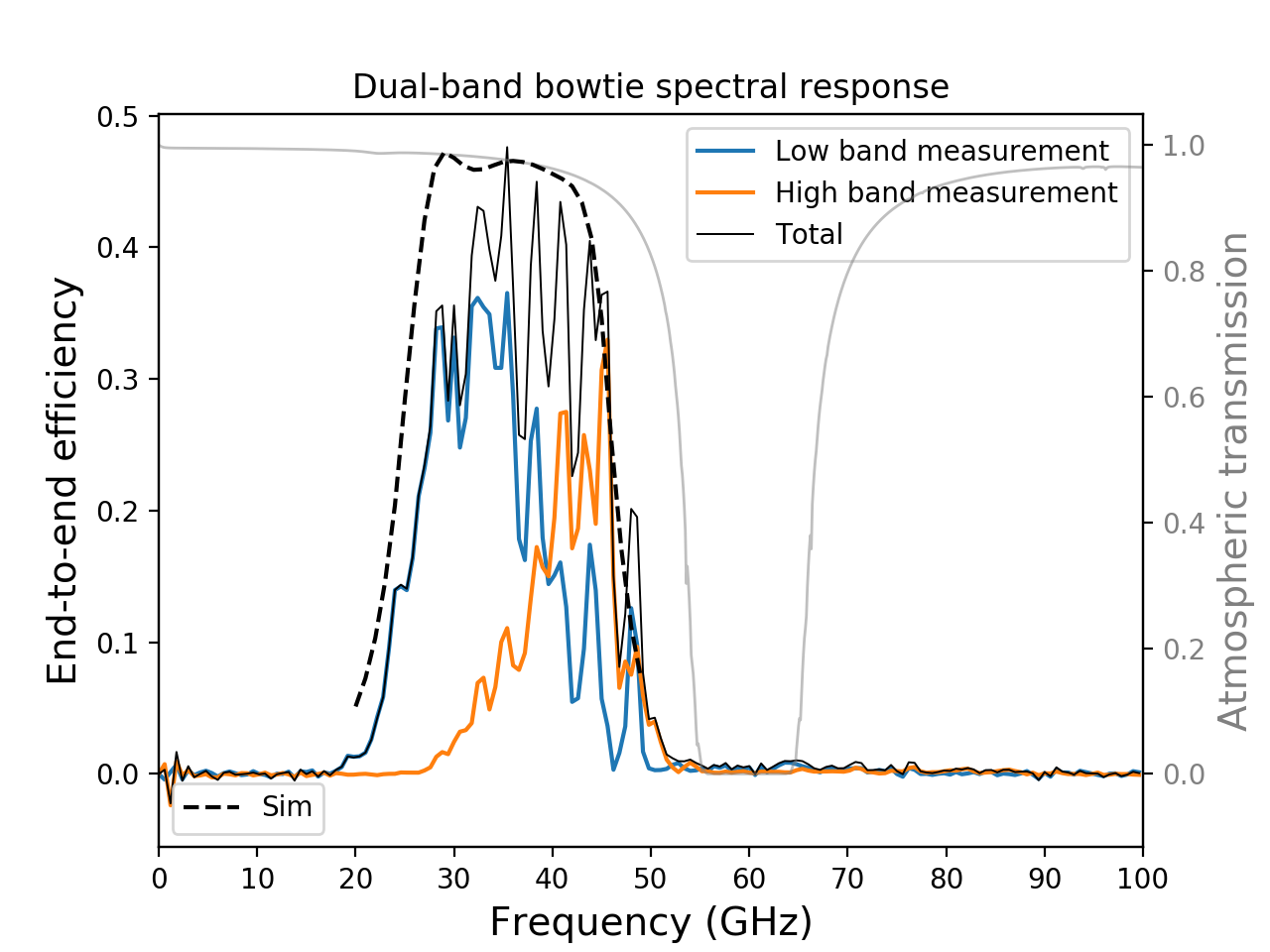} 
\caption{  Comparison between the design spectrum (in the dashed lines) and measured spectrum (in solid lines). The left Figure highlights the simulated spectra, while the right Figure highlights the simulated spectra. 
(Left) Simulations of the antenna transmission spectrum and the intended bands from the on-chip filters. Measured spectra have been scaled vertically for visual comparison. A comparison of simulated and measured values shows that the diplexer shifted up 4\,GHz from intended design values, causing the lower-band to encroach on the bandwidth of the higher-band. This leads to unequal bandwidths between the two bands. 
(Right) An unknown impedance step causes a $\sim$3 GHz ringing.  The atmospheric transmission at the South Pole is shown in gray, with molecular oxygen responsible for the line at 60\,GHz \citep{oxygen_line}. } 
\label{fts_spectrum}
\end{figure*}

 The spectral response of these antennas, $S(\nu)$,  was characterized using a Martin-Puplett Fourier transform spectrometer (FTS). The input light is generated by an HR-10 source submerged in liquid nitrogen for thermal stability of the source.
 First, the mm-wave radiation passes through a wiregrid, selecting for linear polarization. 
 Subsequently, it is directed to a collimating mirror and onto a beam splitter consisting of a wiregrid oriented 45 degrees to the incident polarization.
 The beam is split into two: one path is a fixed rooftop mirror, and the other is a variable rooftop mirror continuously driven by a stepper motor. 
 The two paths interfere when returning to the beam splitter, reflecting off an output wiregrid. This wiregrid is oriented at 45 degrees so that only the desired linear polarization leaves the FTS box, thereby cleaning up any undesired polarized systematics. 
 Lastly, the beam is focused by a HDPE lens designed to illuminate a single pixel in the optical testbed. 
 We do not believe that this source is beam-filling, so the throughput on the single detector is expected to be $A\Omega = f \lambda^2$, where $f$ is a fractional value. However, it's important to note that the FTS source operates in the Rayleigh-Jeans limit: $I(\nu) = k_b T \lambda^{-2}$. Therefore, the reported detector response $S(\nu)$ is equivalent to the response to a source with constant spectral radiance. 
 

 The spectral response of the detectors is defined in the  band center by, 
\begin{equation} 
\nu_0 \equiv \frac{\int \nu S (\nu) \; d \nu }{\int S(\nu) \; d \nu} 
\end{equation}
and bandwidth by, 
\begin{equation} 
\Delta \nu \equiv \frac{ \left( \int S (\nu) \; d \nu \right)^2}{\int (S (\nu))^2 \; d \nu }
\end{equation}
The normalization of our spectra was computed by measuring the optical response of our detectors. This was performed by comparing the optical loading of the detectors under a beam-filling blackbody source at room temperature and at liquid nitrogen temperatures. Therefore, the spectra in Figure \ref{fts_spectrum} represent the end-to-end optical efficiency of the antennas to in-band photons, accounting for losses through the optical elements such as the window and mm-wave filters, as well as electrical losses through impedance mismatches and dielectric loss in the microstrip summing tree.

The left side of Figure \ref{fts_spectrum} shows the simulated antenna bandwidth as the dashed black line, along with simulated on-chip filters defining the two bands. 
The simulated antenna spectrum suggests additional usable bandwidth; however, the oxygen line at 60\,GHz must be rejected using on-chip filters. 
The diplexer continuously distributes power between the two bands. There is spillover between the two bands as the filter is optimized to minimize reflected power between the bands rather than band separation. 
This design choice was made for two reasons: (1) avoiding power loss between the two bands, as there are no atmospheric lines between 30 and 40\,GHz, and (2) opting for a conservative filtering scheme. This design avoids reliance on two electrically interacting on-chip filters, where drift in one would interfere with the other through 3-port interactions.
Ultimately, this spectral spillover reduces our lever arm in measuring $\beta_s$, as the effective frequency centers of both bands are closer in frequency space.

The right side of Figure \ref{fts_spectrum} shows the full-focal plane averaged spectrum of both bands in blue and orange. The sum of the two bands is the solid black line, indicating that physical measurements show a consistent bandwidth between measurements and our simulated expectations. 
Tabulated bandcenters and widths are in Table \ref{sum_optical}. High fractional bandwidth is achieved with both the high or low bands having substantially larger fractional bandwidth than the traditional slot design \citep{cheng}. 

Power is split between the two contiguous channels, indicating that the diplexer performs as intended. The upper and lower-band edges agree with our simulated results.
The diplexer appears to have shifted from design frequency by 10\%, which causes more spillover from the lower band to the upper band than intended.

The achieved optical efficiency is modest, with end-to-end measurements of 20-30\% optical efficiency. 
A portion of these quoted lower optical efficiencies occurs due to spectral spillover; a sum of bands shows 35\% efficiency in the passband. 
While this efficiency remains lower than the monochromatic slot design, these detectors remain competitive with similar optical responsivity due to larger overall bandwidth. 

Additionally, there is an observed $\sim$3 GHz ringing in the frequency bands, suggesting the presence of an impedance mismatch associated with this length scale or one of its harmonics.
The effective dielectric constant of microstrip modes in a 0.3$\,\mu m$ SiO$_2$ is approximately $\epsilon_{\text{MS, eff}} \sim 3.4$, which results in a corresponding wavelength of $\sim 54\,$mm. Unfortunately, many possible junctions within the summing tree have this length scale, making it challenging to identify the source of this impedance mismatch. 

In future iterations of this design, we would adjust the diplexer to shift the frequency down so that the two bands have a more balanced distribution. Alternatively, we can employ a different filtering scheme to achieve a sharper separation between the two bands at the expense of some loss of photons between the two bands. A preliminary design has been explored in Appendix \ref{appendix:new_diplexer}.
Additionally, more investigative work must be done to identify the source of impedance mismatches in the microstrip network to identify the source of reflections. 
Highlighting that this design meets the baseline requirement within BICEP Array to achieve its science goals, there remains significant potential for enhancing the end-to-end efficiency of these detectors to meet simulated expectations.  
\FloatBarrier
\subsection{Far-Field Beam Patterns}
\label{sec:beams}
\begin{figure*} 
\begin{minipage}{0.33\textwidth}
\includegraphics[width = \textwidth]{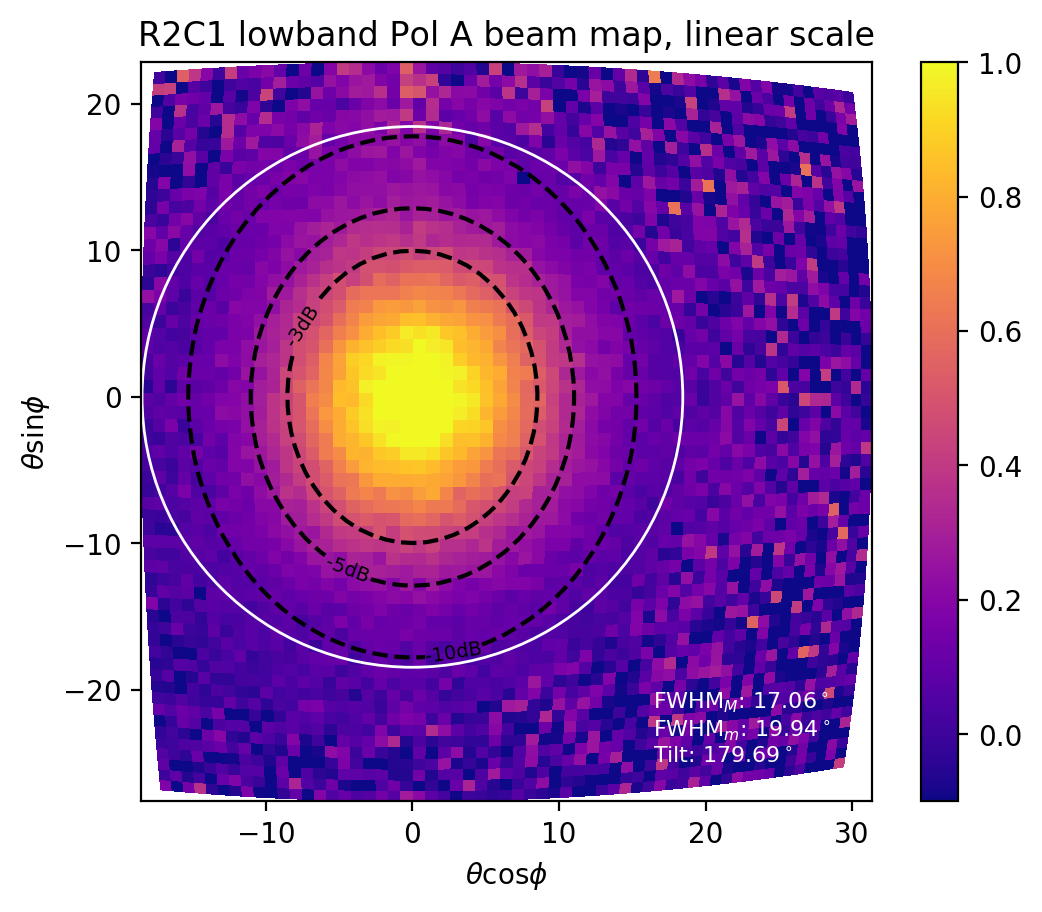}
\end{minipage}\begin{minipage}{0.33\textwidth}
\includegraphics[width =\textwidth]{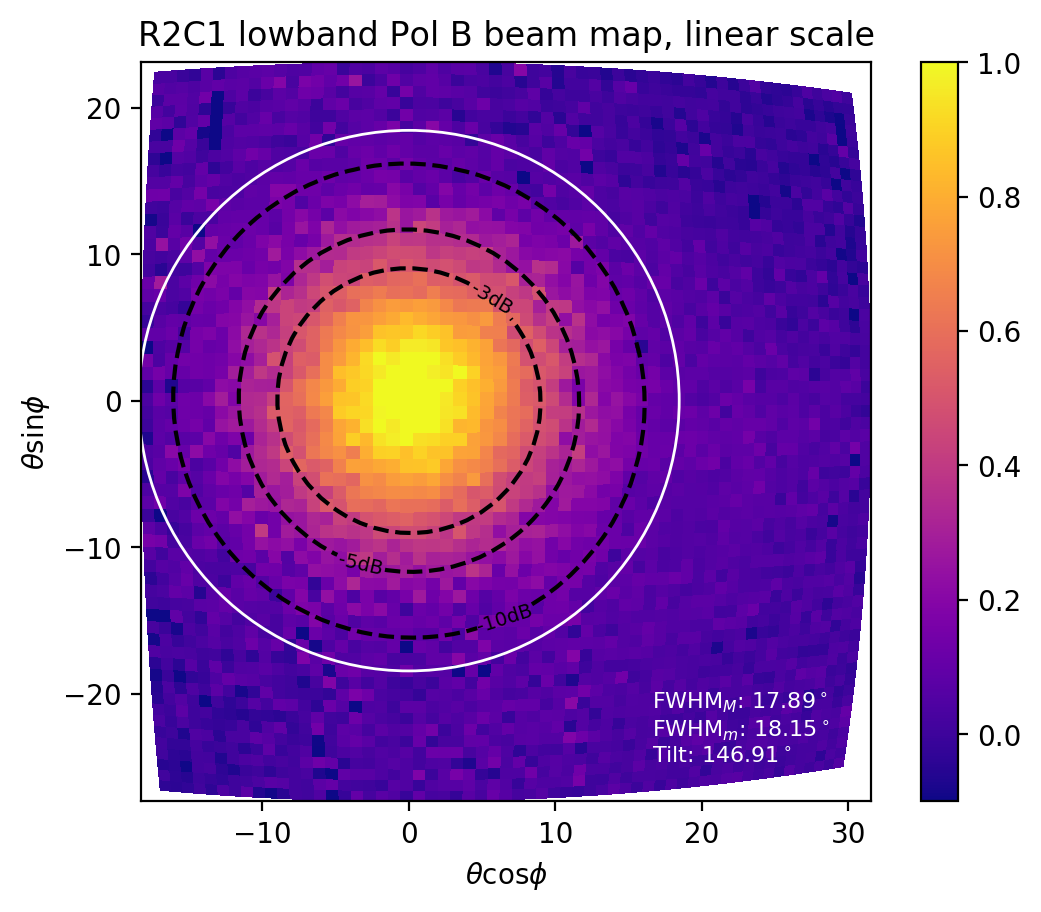} 
\end{minipage} \begin{minipage}{0.33\textwidth}
\includegraphics[width =\textwidth]{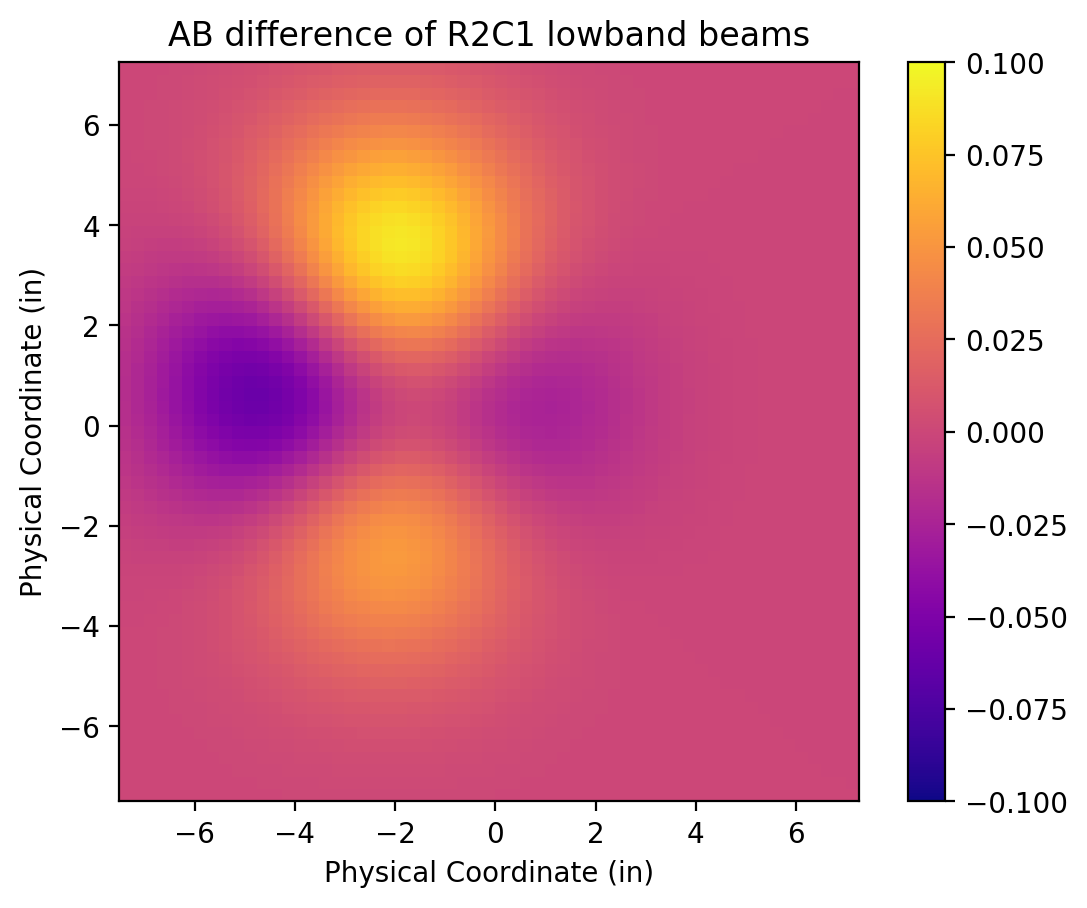} 
\end{minipage} 
\begin{minipage}{0.33\textwidth}
\includegraphics[width = \textwidth]{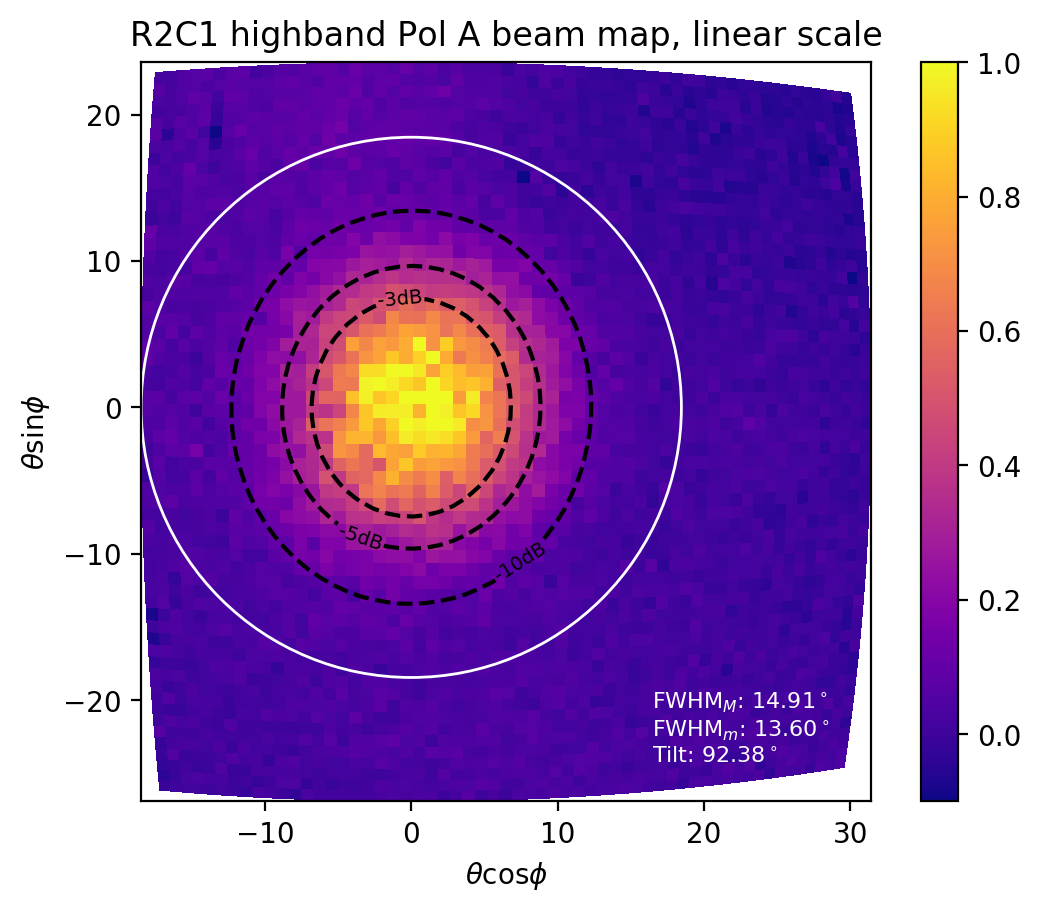}
\end{minipage}\begin{minipage}{0.33\textwidth}
\includegraphics[width =\textwidth]{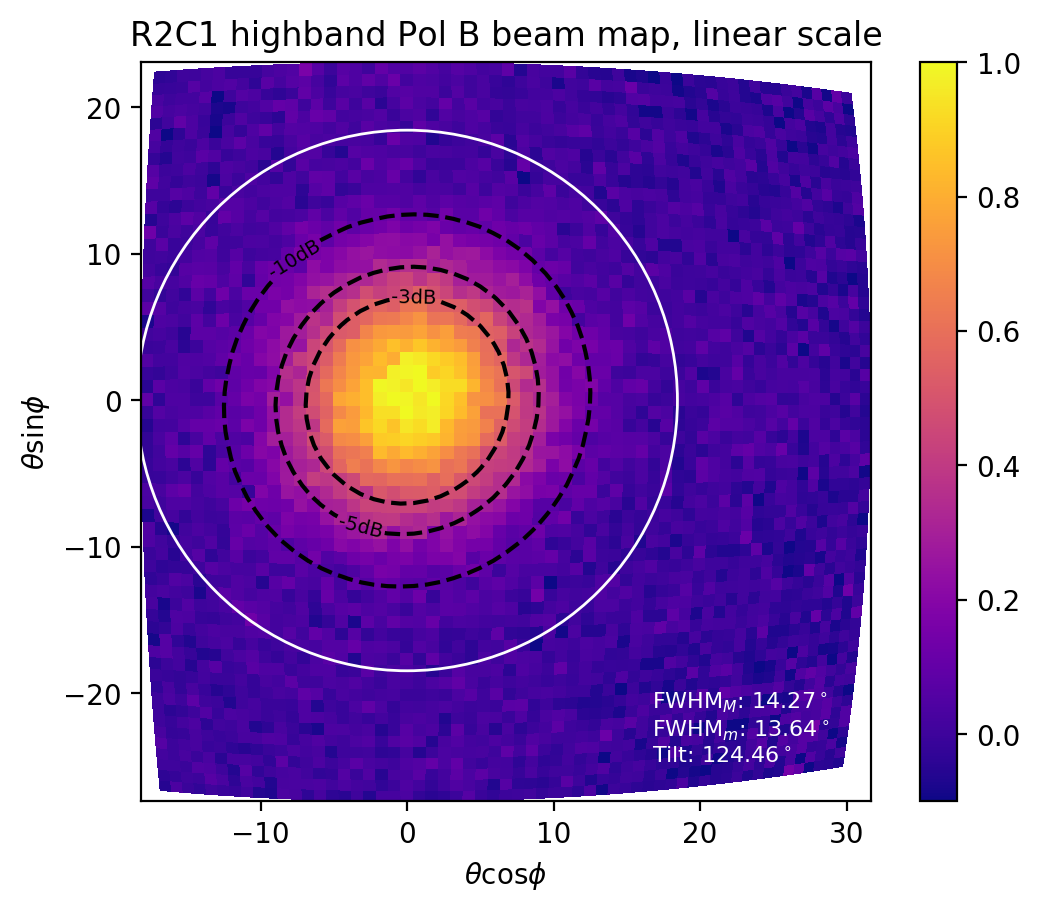}
\end{minipage}\begin{minipage}{0.33\textwidth}
\includegraphics[width =\textwidth]{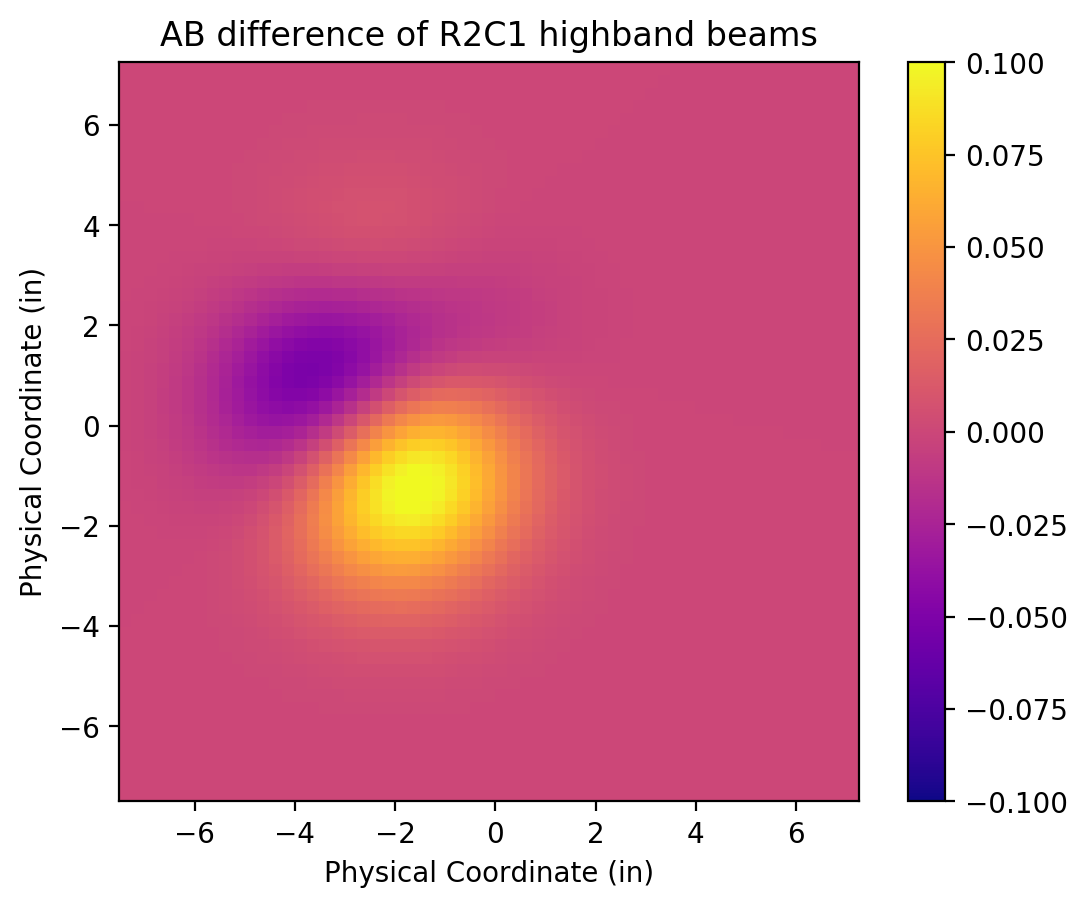}
\end{minipage}

\caption{Farfield beam map measurements of a representative low-band detector pair (upper row) and a high-band detector pair (lower row). These four detectors are low and high-band A/B polarization pairs from an edge-pixel. All plots show the beam map and a F2D Gaussian fit with -3, -5, -10dB black dashed contour line overlayed on top. The white line overlays the optical footprint of a $f/1.5$ system. Polarized frame edge affects the Pol. A antennas and causes an ellipticity of the main beam. Pol B is unaffected by the frame. The 2D Gaussians are differenced in the last column, showing A-B differences. The two peak-to-peak differences in the dipole moment for the low-band are 9\% and -6\%, and for the high-band are 10\% and -5\%. This detector shows typical performance. Peak-to-peak variations as low as 10\% has been measured on edge-pixels, which is near the $\sim 8\%$ theoretical limit. } 
\label{fig:beam_map} 
\end{figure*}

  \begin{table*}[t] 
  \caption{Summary Statistics of the Optical Performance of the Diplexed Focal Plane. A shift in the diplexer central frequency has caused spillover of the lower-band bandpass, enroaching on the bandwidth of the upper-band. The beam-ellipticity statistics is driven by undesired polarized frame and edge-pixels electromagnetic interaction, which make up the majority of the pixels on a $4\times4$ focal plane. Notably, center pixels and edge pixels with the orthogonal polarization to the frame, show significant improvements to their beam ellipticities.} 
  \begin{tabular}{r lll@{\hskip 0.3in} lll} \toprule
  &  \multicolumn{3}{c}{Spectra} &  \multicolumn{3}{c}{Beams} \\
  & $\nu_0$ & $\Delta \nu $ & $\frac{\nu_0}{\Delta \nu}$ & FWHM$_M (^\circ) $ & FWHM$_m (^\circ)$ & $\Delta(M - m) ({}^\circ)$ \\ \midrule
  Low-band & 33.7\,GHz & 21.2\,GHz & 63\% & 18.5 $\pm$ 1.3 & $16.9 \pm 0.9 $ & $1.6 \pm 1.2$ \\
  High-band & 41.5\,GHz & 15.4\,GHz & 37\% & 15.1 $\pm 1.6$ & 14.1 $\pm 1.4$  & $0.8 \pm 0.4 $\\ \bottomrule
  \end{tabular} 
  \label{sum_optical}
  \end{table*} 
The antenna beam of a phased antenna array is, in principle, highly tunable. The configuration of the radiators, along with the amplitude and phase applied to each antenna element, collectively determine the beam shape. 
The gain of an antenna array is modeled as the following: 
\begin{equation} 
G_{\text{array}}(\theta, \phi) = G_0(\theta, \phi) |AF(\theta, \phi)|^2
\label{eq:antenna_array} 
\end{equation}
where $G_0$ is the gain of an individual element, and $AF$ is short for the array factor. The array factor is the discrete Fourier transform of the antenna elements. Generically, the array factor can be expressed, 
$AF(\theta, \phi) = \sum_{n,m} A_{nm}\exp \left(i  \frac{2 \pi}{\lambda} \hat{r}\cdot \vec{x}(n,m) \right)$, where $\hat{r} = \sin \theta \cos \phi \hat{x} + \sin \theta \sin \phi \hat{y}$ with $\theta,\phi$ being the polar and azimuthal angle respectively, $A_{nm}$ is the excitation amplitude of each radiator, and $\vec{x}(n,m)$ is the location of each radiator. 

The number of sub-radiators was chosen to match the telescope optics of f/1.5 in BICEP Array \citep{howard}, but in principle, this can be tunable to any optical system. 
For a single polarization, a pixel comprises of a 12x12 square grid of resonant bowtie antenna pairs. This parameterization is convenient as it allows us to analytically separate the array factor into the product of two array factors.  

\begin{align} 
|AF(\theta, \phi)| &= \frac{\sin \left(\frac{2 \pi M a}{\lambda} \sin \theta \cos \phi \right)\sin \left(\frac{2 \pi M a}{\lambda} \sin \theta \sin \phi \right)}{M^2 \sin \left( \frac{2 \pi a}{\lambda} \sin \theta \cos \phi\right)\sin \left( \frac{2 \pi a}{\lambda} \sin \theta \sin \phi\right)} \times \nonumber \\ 
& \qquad \qquad \cos \left(  \frac{2 \pi}{\lambda}  \frac{a}{2} \sin \theta (\cos \phi \pm \sin \phi)\right)
\end{align}

The first term represents the $M = 12$ square array factor. The second term represents the two-element antenna pair, with the $\pm$ for the A and B polarizations respectively. The sign choice originates from antenna pairs being located at quadrants 1 and 3 for polarization A and quadrants 2 and 4 for polarization B. 
This beam model implies an inherent A-B mismatch between the two polarization pairs, arising from the two-element component portion of the array factor. 
Indeed, the A-B mismatch for this component increases with larger polar angles from the antenna boresight. 
However, the square array factor concentrates the antenna beam and minimizes this source of A-B mismatch.
A calculation shows, for a 12$\times$12 array, this pairwise array factor contributes only a 0.7\% peak-to-peak A-B mismatch. 
The intrinsic antenna element beam, $G_0$, is determined  through HFSS simulations and contributes an additional 2\%, leading to a theoretical limit of a 2.7\% peak-to-peak A-B mismatch. 

Previous antenna arrays have used an $8\times$8 array. This is expected to have a 1.6\% peak-to-peak A-B mismatch purely from the pairwise array factor, and a 5\% peak-to-peak A-B mismatch when accounting for individual beams. 
The increased number of elements for this antenna array design suppresses differential ellipiticity and leads to overall better beam performance. 

 Far-field beams were characterized by an in-house beam mapper, which consists of a Thorlabs MC2000 optical chopper: a variably controlled (2-20Hz) thermally chopped source which is mounted on an X-Y translation stage. The thermal source is a heated ceramic, reaching temperatures of several hundreds of degrees Celsius, and is chopped with respect to a reflective room-temperature blade. The reflected beams terminate within the cold optics of the testbed. 
 To prevent multiple reflections in our measurement setup, we took care to blacken the beam mapper facing the detectors with Eccosorb HR-10.
 Since there are no focusing optics in the testbed cryostat, the chopped source was located in the far field of the detector array. 

 We demodulate the time-ordered data using the chopper optical encoder as reference, so that only variations at the chop frequency and in phase is interpreted as signal. This reference signal was read out using the same readout electronics employed for the detectors. 

Out-of-band photons, rather than being captured by the antennas and filtered out by the on-chip filters, can inadvertently directly illuminate the bolometer island. 
The exact mechanism for this response is not fully understood, but a plausible mechanism is that it caused by potential differences in the bolometer island relative to the ground plane, which drive currents in the resistive termination and heat the bolometer island \citep{cheng_thesis}.

In order to control for the baseline response of the detectors, we measured the direct island response using out-of-band, high-frequency photons. This response is mapped out and subtracted from our nominal beam maps. 
These data were acquired by filtering the source with a specially designed thickgrill filter, a one\,cm thick metallic plate with a dense array of circular waveguide apertures. 
The filter is designed to permit only power above the waveguide cutoff, 61\,GHz, allowing illumination solely through direct island stimulation.

Figure \ref{fig:beam_map} shows the resulting beam-map, corrected for direct island illumination. Each of the beam-maps are fitted to a 2-D Gaussian with full-width at half maximum (FWHM) in the plot inset and -3dB, -5dB, and -10dB contours are plotted on top. We also include the $f/1.5$ optics on top of the beam-map. 
The theoretical expectations for 12x12 pairs of antennas, convolved with the beam of a single element bowtie antenna, match well with the measured results. For a $\delta$-function response at 35\,GHz and 45\,GHz, we compute a FWHM of 19.6$^\circ$ deg and 15.8$^\circ$ deg respectively. Real beam-maps are an average over the entire band. 

On the right most panel of Figure \ref{fig:beam_map}, we show A-B differences for this representative edge pixel. 
A simulated finite antenna array, with no edge-effects, is expected to have $\sim 3\%$ peak-to-peak differences in a quadrupole pattern. 

Edge effects caused by a solid frame are expected to cause strong differential pointing and result in peak-to-peak variations of 30-40\%, depending on the precise distance between the frame and the antenna array \citep{Ahmed_thesis}. 
An optimal spacing and corrugated frame is expected to achieve polarized beam residuals of less than $10\%$ peak-to-peak.
Our measurements show that a representative edge pixel has 15\% peak-to-peak beam residuals. This is similar to beam maps from center-pixels, suggesting that the polarized beam residuals are not dominated by electromagnetic interactions with the corrugated frame.

\section{Conclusion} 

At any frequency, dichroic detectors offer twice the number of detectors in the same focal plane footprint, allowing a $\sqrt{2}$ improvement in noise-equivalent temperature (NETs). However, at these lower CMB observation frequencies, pixel sizes must scale accordingly with the wavelengths, resulting in naturally smaller values for $N$. The NETs show steeper improvement in this regime, and therefore, there is a larger absolute benefit from increasing $N$ at these lower frequencies. 

We have developed a compact, linearly polarized antenna with a broad first resonance. This first resonance can effectively be divided into two distinct frequency bands, operating at 30/40\,GHz.  

This antenna is compact and well-suited for integration into BICEP's phased antenna array architecture. The detectors are entirely planar and fully lithographed in thin films without the need for any focusing optics. The top-hat illumination maximizes the directivity of the antenna beam within the smallest area. 
It exhibits beam characteristics on par with the slot antenna arrays, making it a suitable choice for CMB polarimetry. Overall, the antenna array architecture enables the highest pixel density for a fixed focal plane area. 

This detector array deployed in the inaugural season of BICEP Array, and is the first demonstration of a dichroic planar array system ever used for CMB measurements at the South Pole \citep{Ahmed_thesis}. 
This work enhances the sensitivity of BICEP array towards low-frequency foregrounds. The dual-frequency measurement enables simultaneous measurement of both the synchrotron amplitude and its spectral index, offering a more powerful constraint on this source of foreground for CMB polarimetry. Further on-sky synchrotron data, recorded by this detector, is being analyzed for a potential publication soon. 

Immediate future work would be to improve the optical efficiency of the detectors by identifying the source of impedance mismatch in the microstrip summing tree. 
Furthermore, the potential for higher frequency focal planes (220-270\,GHz) would allow for ever more precise constraints of thermal dust emission. 
However, the scalability of the current design poses certain challenges. 
While the antenna elements themselves would scale proportional to wavelength, the microwave feed network encounters non-trivial scaling issues. 
The impedance of microstrip lines and desired spacing between these lines may not necessarily decrease with higher frequencies. 
The existing design, already constrained by space at these frequencies, poses a challenge in scaling down the antenna elements while maintaining similar microstrip trace widths. These challenges have been preliminary addressed in this publication \citep{Ahmed_thesis} for two CMB observation bands at 90/150\, GHz, simultaneously. Additionally, dielectric loss is larger at higher microwave frequencies leading to expected degraded performance. These challenges underscore the considerable efforts required to extending the design to higher frequencies. Addressing these issues would allow us to place more detectors in the sky which is especially important at higher frequencies where detectors are more easily photon-background limited. 
\begin{acknowledgements}

This research was carried out (in part) at the Jet Propulsion Laboratory, California Institute of Technology, under a contract with the National Aeronautics and Space Administration and funded through JPL’s Strategic University Research Partnerships (SURP) program. 
This publication is also supported by the National Aeronautics and Space Administration grant No. NNX17AC55G.

\end{acknowledgements}
\FloatBarrier
\appendix
%


\section{Improving Diplexer Separation}
\label{appendix:new_diplexer}
\begin{figure*} 
\begin{minipage}{0.49\textwidth}

 \begin{tabular}{rccccc} \toprule  
 & $L_1$ & $C_1$ & $L_2$ & $C_2$ & $C_3$  \\ \midrule 
 Bandpass (30GHz) & 3.54 & 0.78  & 4.45 & 0.78 & 1.14 \\
 Bandpass (40GHz) & 4.42 & 0.58 & 5.35 & 0.58 & 1.15 \\ \bottomrule 
 \end{tabular} 
\end{minipage}
\begin{minipage}{0.49\textwidth}
 \begin{circuitikz}[scale = 0.7, transform shape, /tikz/circuitikz/bipoles/length = 0.9cm] 
 
 \draw(-1.5,3) node[below]{to Antennas} to [short, o-](0,3) to[short](0,2); 
 \draw(0,3) to[short](0,4); 
 
 \draw(0,4) to[C, l = $C_1$](1,4) to[L, l = $L_1$](2,4) to[C, l = $C_1$](3,4);  
 \draw(3,4) to[short](3,3.75) to[C, l = $C_3$](3, 2.75) node[ground]{}; 
 \draw(3,4) to[C, l = $C_2$](4,4) to[L, l = $L_2$](5,4) to[C, l = $C_2$](6,4); 
 \draw(6,4) to[short](6,3.75) to[C, l = $C_3$](6,2.75) node[ground]{}; 
 
 \draw(6,4) to[C, l = $C_1$](7,4) to[L, l = $L_1$](8,4) to[C, l = $C_1$, -o](9,4);  
 \draw(9, 3.5) node{High pass};

 \draw(0,2) to[C, l = $C_1$](1,2) to[L, l = $L_1$](2,2) to[C, l = $C_1$](3,2);  
 \draw(3,2) to[short](3,1.75) to[C, l = $C_3$](3, 0.75) node[ground]{}; 
 \draw(3,2) to[C, l = $C_2$](4,2) to[L, l = $L_2$](5,2) to[C, l = $C_2$](6,2); 
 \draw(6,2) to[short](6,1.75) to[C, l = $C_3$](6,0.75) node[ground]{}; 
 
 \draw(6,2) to[C, l = $C_1$](7,2) to[L, l = $L_1$](8,2) to[C, l = $C_1$, -o](9,2);  
 \draw(9, 1.5) node{Low pass};

 \end{circuitikz} 
\end{minipage}
\caption{A diplexer composed of two bandpass filters. Each bandpass filter is composed of a three-pole LC resonator circuit. The design table for the circuit network is tabulated on the left. To scale from design table to physical values: inductor values are scaled by $\omega_0^{-1} Z_0$  and capacitor values are scaled by $(\omega_0 Z_0)^{-1}$ where $\omega_0 = 2\pi f_0$ is the desired central frequency of the bandpass filters ($30$ and $40\,$GHz), and $Z_0$ is the port impedance. For this design, $Z_0 = 10\,\Omega$ was lowered to achieve smaller and more physically realizable inductor values. Simulations are then performed to convert the electrical inductance and capacitance to a lithographic element.} 
\label{fig:diplexer_dual_bandpass}
\end{figure*}
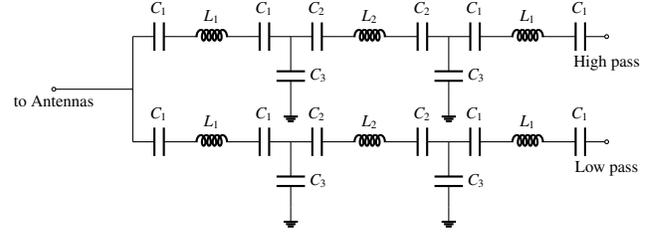

The diplexer design in Section \ref{sec:filtering} is optimized to minimize reflections in band but exhibits significant spillover between the two bands. The band overlap diminishes our ability to measure the synchrotron SED slope. 
In a separate fabrication run, we optimized the diplexer for band separation. 

We designed a diplexer shown in Figure \ref{fig:diplexer_dual_bandpass}, consisting of two parallel bandpass filters, each centered at 30 and 40\,GHz with a 10\,GHz bandwidth. 
The architecture is identical to previously designed bandpasses: a three-pole LC-tank joined with shunt capacitors acting as impedance inverters. 

In order to minimize electrical interactions between the two bands, we found it advantageous to have a steep cutoff for the bandpass filters. When the bandpass filter is out of band, the resistance drops to zero and the reactance diverges. A sharper bandpass results in less interaction between the two filters. 
We found it advantageous to use a diplexer with 0.5dB ripples or have variations of up to 12\% in the passband. 

\begin{figure*} 
\centering
\begin{minipage}{0.4\textwidth}
\includegraphics[width = \textwidth]{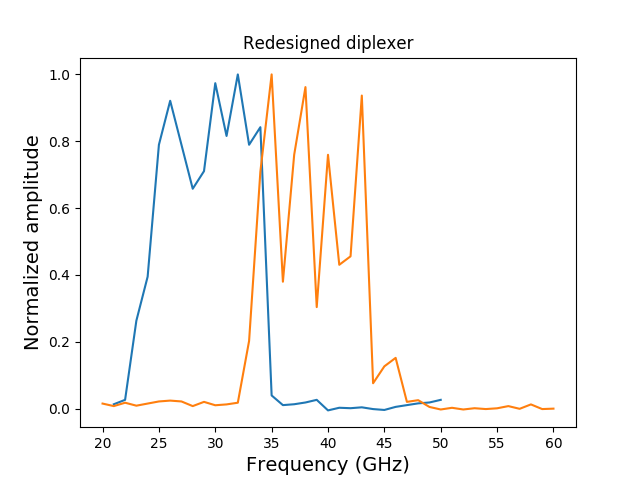}
\end{minipage}
\begin{minipage}{0.55\textwidth}
\includegraphics[width = \textwidth]{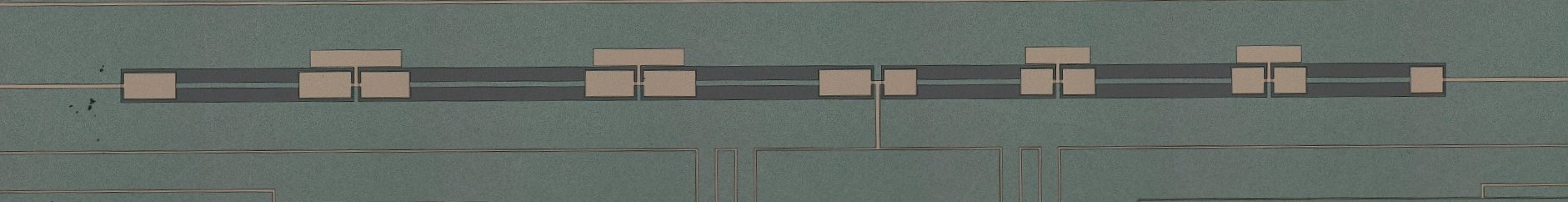}
\end{minipage}
\caption{(Left) Measured spectra of the double-bandpass diplexer from a pair of characteristic pixels. We have a stronger band definition between the two bands. However, an impedance mismatch degrades the performance of these pixels. The photograph on the right shows the newly designed diplexer consisting of two bandpass filters. Each bandpass comprises a 3-pole series LC resonator circuit joined by shunt capacitors.  } 
\label{fig:new_diplexer_spectra}
\end{figure*}

Achieving this steep cutoff required a substantial increase in inductor values. However, achieving these inductor values lithographically would be challenging at our operational frequencies. 
Instead, we found it advantageous to decrease the diplexer's operating impedance, $Z_0$, from 25\,$\Omega$ to 10\,$\Omega$ to decrease inductor values while keeping the capacitors within achievable values. 
Additionally, unlike our previous bandpass filter, we did not need to perform a $Y$ to $\pi$ transformation for the capacitive network. Therefore, Figure \ref{fig:diplexer_dual_bandpass} shows the circuit topography optimized for our design. This is then translated to lithography in the right side of the Figure \ref{fig:new_diplexer_spectra}

The measured spectra in Figure \ref{fig:new_diplexer_spectra} shows that this diplexer scheme achieves sharper band definitions between the two pixels. 
However, strong ringing in the passband degrades pixel performance, resulting in lower optical response. This design requires further investigation to identify the impedance mismatch's source.

\FloatBarrier
\section{Procedure to Convert Lumped elements to Lithography with Sonnet}
\label{appendix:sonnet}
\begin{figure*} 
 \begin{minipage}{0.29\textwidth}
 \centering
\begin{circuitikz}[scale = 0.45]
\draw(0,0) to[short, o-](2,0) to[C,l = $C$](3,0) to[L,l = $L$](5,0) to[C,l = $C$](6,0) to[short, -o](8,0); 
\draw(1, -2) node[ground]{} to[C, l = $C_p$] (1,-0.7) to (1,0); 
\draw(7,0) -- (7,-0.7) to[C, l = $C_p$](7,-2) node[ground]{};
\end{circuitikz}
\end{minipage}
\begin{minipage}{0.35\textwidth}
 \includegraphics[width = \textwidth]{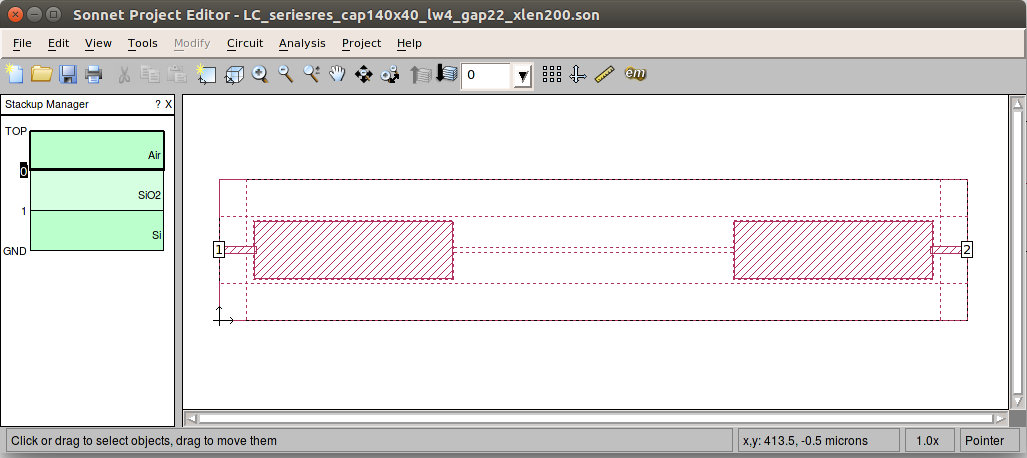}
 \end{minipage}
 \begin{minipage}{0.35\textwidth}
\includegraphics[width = \textwidth]{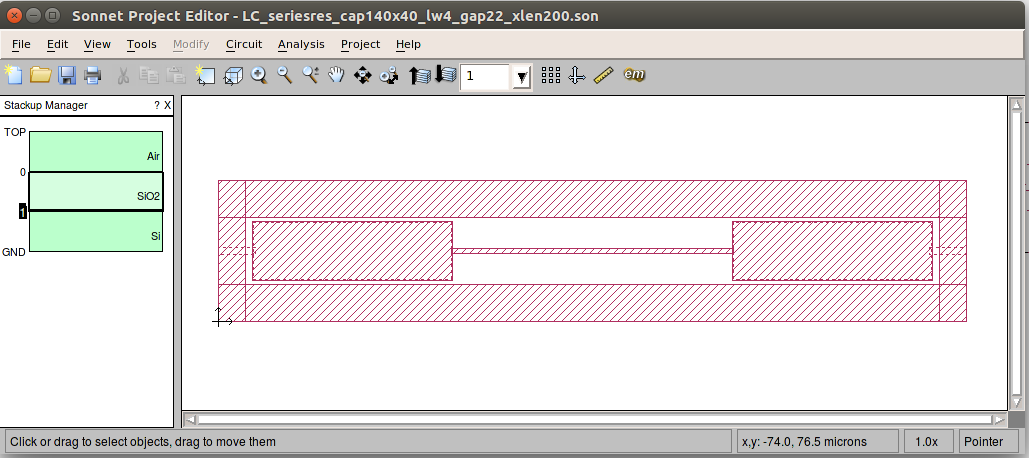} 
\end{minipage}
\caption{The leftmost figure shows an equivalent circuit model to the LC resonator shown in the two figures on the right: a series LC resonator with parasitic capacitances to ground. The middle image shows the Sonnet simulation from the top microstrip trace layer, and the rightmost image shows the Sonnet simulation from the ground plane layer. } 
\label{fig:sonnet}
 \end{figure*}

 This appendix documents our procedure for converting lumped element to a lithographic design. The fundamental building block of a bandpass filter is the LC resonator, consisting of a series inductor and series capacitor as shown in Figure \ref{fig:sonnet}. 
 
 The output of a Sonnet simulation is an S-matrix of the two-port network. These can be converted to an impedance matrix,

\begin{align} 
z_{11} &= Z_0 \frac{(1 + s_{11})(1 - s_{22}) + s_{12} s_{21}}{(1 - s_{11})(1 - s_{22}) - s_{12}s_{21}}\\
z_{12} &= Z_0 \frac{2 s_{12}}{(1 - s_{11})(1 - s_{22}) - s_{12}s_{21}}\\
z_{21} &= Z_0 \frac{2 s_{21}}{(1 - s_{11})(1 - s_{22}) - s_{12}s_{21}}\\
z_{22} &= Z_0 \frac{(1 - s_{11})(1 + s_{22}) + s_{12} s_{21}}{(1 - s_{11})(1 - s_{22}) - s_{12}s_{21}}
\end{align}

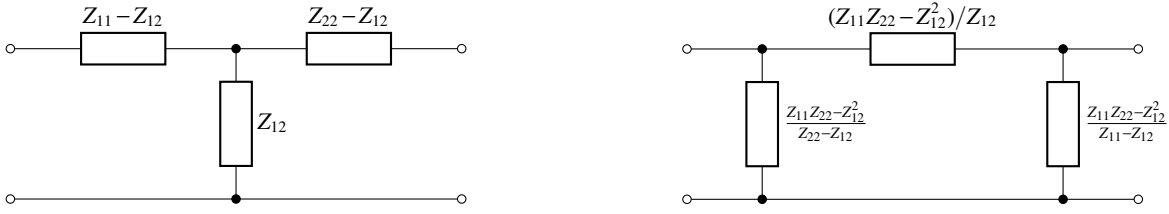
\begin{figure*} 
\begin{minipage}{0.49\textwidth}
\begin{circuitikz}[european] 
\draw(0,0) to[R, l = $Z_{11} - Z_{12}$, o-](3,0) to [R, l = $Z_{22} - Z_{12}$, -o](6,0);
\draw(3,0) to[R, l = $Z_{12}$, *-*](3, -2); 
\draw(0, -2) to[short, o-o](6,-2); 
\end{circuitikz}
\label{}
\end{minipage} 
\begin{minipage}{0.49\textwidth}
\begin{circuitikz}[european] 
\draw(0,0) to[R, l = $ (Z_{11} Z_{22} - Z_{12}^2)/Z_{12}$, o-o](6,0); 
\draw(1,0) to[R, l = $ \frac{Z_{11} Z_{22} - Z_{12}^2}{Z_{22} - Z_{12}}$, *-*](1, -2); 
\draw(5,0) to[R, l = $\frac{Z_{11}Z_{22}- Z_{12}^2}{Z_{11} - Z_{12}}$, *-*](5, -2); 
\draw(0, -2) to[short, o-o](6,-2); 
\end{circuitikz}
\label{fig:circ_rep}
\end{minipage} 
\caption{(Left) A generic two-port representation with impedance parameters (Right) An equivalent representation.  } 
\label{fig:circ_rep_all}

\end{figure*}

The equivalent circuit for the $Z$-matrix of any reciprocal two-port network is shown in Figure \ref{fig:circ_rep_all}. 
Therefore, a generic series impedance from a two-port network can be calculated by the impedance parameters as demonstrated in Figure \ref{fig:circ_rep_all}, 
\begin{equation} 
Z =  \frac{Z_{11} Z_{22} - Z_{12}^2}{Z_{12}} .
\label{genericz}
\end{equation}
Additionally, the circuit model of an LC resonator, shown in Figure \ref{fig:sonnet}, has the impedance
\begin{equation} 
Z = i \omega L + \frac{2}{i\omega C} = i \left( \omega L- \frac{2}{\omega C} \right).
\end{equation}  
The derivative of this expression gives us an additional constraint equation, 
\begin{align} 
\frac{\partial Z}{\partial f} &= 2 \pi \frac{\partial Z}{\partial \omega}  = \frac{2 \pi }{\omega } i \left(  \omega L  + \frac{2}{\omega C} \right)
\end{align}
Expressing these into a system of equations,
\begin{equation} 
\begin{pmatrix} Z \\ \frac{\omega}{2 \pi} \frac{\partial Z}{\partial f} \end{pmatrix} =i \begin{pmatrix} 1 & -2 \\ 1 & 2 \end{pmatrix} \begin{pmatrix} \omega L \\ \frac{1}{\omega C} \end{pmatrix} 
\end{equation} 
and the corresponding inverse, 
\begin{equation} 
\begin{pmatrix} \omega L \\ \frac{1}{\omega C} \end{pmatrix}  = - \frac{i}{2} \begin{pmatrix} 1 & 1 \\ -\frac{1}{2} & \frac{1}{2}
\end{pmatrix} \begin{pmatrix} Z \\ \frac{\omega}{2 \pi} \frac{\partial Z}{\partial f} \end{pmatrix}.
\end{equation}
This allows us to solve for the inductance and capacitance, where $Z$ determined by equation \ref{genericz}. 
\begin{align} 
L &= \frac{1}{2 \omega}  \left( Im[Z]  + \frac{\omega}{2 \pi } Im \left[ \frac{\partial Z}{\partial f}  \right] \right) \\ 
C &= \frac{1}{4 \omega} \left(- Im[Z] +  \frac{\omega}{2 \pi} Im \left[ \frac{\partial Z}{\partial f } \right] \right)^{-1}
\end{align} 

Additionally, the parasitic capacitance is, 
\begin{equation}
C_p = \frac{1}{i \omega} \frac{Z_{22} - Z_{11}}{Z_{11} Z_{22} - Z_{12}^2} = \frac{1}{\omega} \frac{Im[Z_{22}]  - Im[Z_{11}]}{Z_{11} Z_{22} - Z_{12}^2} 
\end{equation}

To develop a lithographed filter, we first compile a comprehensive library of LC resonator simulations for a fixed dielectric stack. 
We use the niobium metal model in Sonnet, which has zero resistance in DC and AC, and accounts for kinetic inductance with $L_s = 0.11\,$pH/sq.
We conduct simulations with various lengths of capacitors and inductors, systematically tabulating the lumped element equivalent models, including the frequency dispersion of both elements. 
Subsequently, when aiming for a specific lumped element model, we interpolate the precise geometric parameters necessary for achieving the desired filter characteristics. 
Then, a realized lithographic filter is fully simulated and tweaked if necessary. 
This iterative process allows us to fine-tune and optimize the lithographic design to meet specific performance criteria and minimize frequency dispersion. 
\FloatBarrier

\section{Aperture Efficiencies of Antenna Arrays}

\label{appendix:aperture_eff}
\begin{figure*} 
\begin{minipage}{0.5\textwidth}
\includegraphics[width = \textwidth]{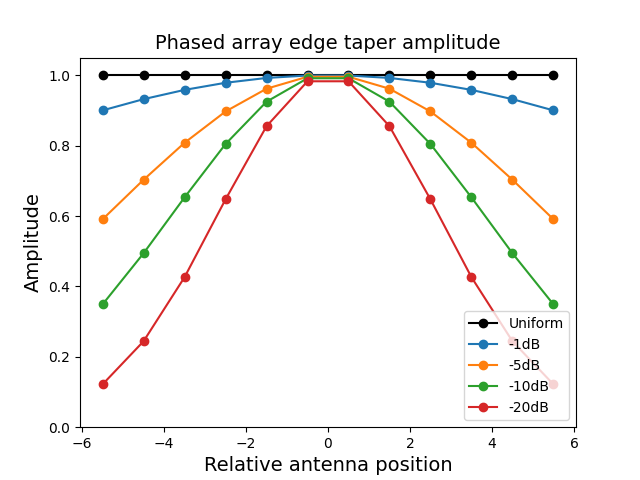}
\end{minipage}
\begin{minipage}{0.5\textwidth}
\includegraphics[width = \textwidth]{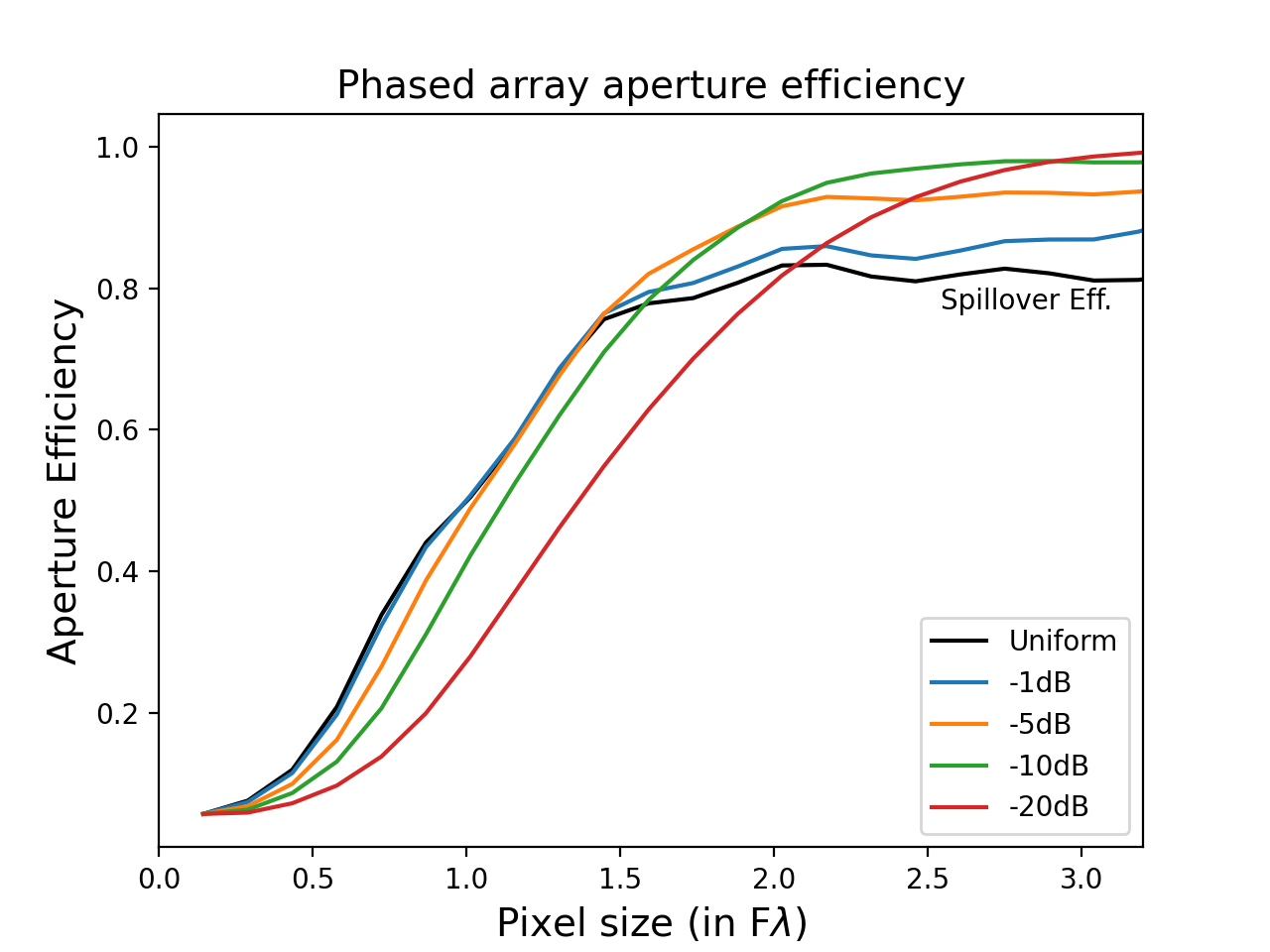}
\end{minipage}
\caption{(Left) Illumination of an example 12-element antenna array with different edge tapers. (Right) Calculation of the aperture efficiency of a $N\times N$ antenna array with different edge tapers. The dashed lines show a uniformly illuminated array's illumination and spillover efficiencies. The antenna array presented in this publication is uniformly illuminated with a pixel size of $\sim1.5,2.0F\lambda$ for the two bands, respectively.} 
\label{fig:aper_eff_edgetaper}
\end{figure*}

Aperture efficiency is a measure of how effectively an optical system captures and utilizes incoming radiation. 
For an extended source like the CMB, aperture efficiency is determined entirely by the spillover efficiency, $\epsilon_s$: \citep{quasioptical_systems} 
\begin{equation}
\epsilon_{\text{aperture efficiency}} = \epsilon_{s}\end{equation}. 
In this section, we assume the antenna array consists of an isotropic radiator, $G_0(\theta, \phi) = 1$, constructing the array beam entirely through the array factor. For a simple lens system with $f\slash \#$, the field of view is extends between 0 and $\theta_0 =\arctan((2f\slash \#)^{-1})$. Consequently, the spillover efficiency represents the fractional power contained within a specified field of view:

\begin{equation} 
\epsilon_{s} =
\frac{\int_0^{2 \pi} d\phi \int_0^{\theta_0} d\theta\; |AF(\theta, \phi)|^2 \sin \theta }{\int_0^{2 \pi} d\phi \int_0^{\pi/2} d\theta\; |AF(\theta, \phi)|^2 \sin \theta }
\end{equation}

where $AF(\theta,\phi)$ is the array factor and $|AF|^2$ is the radiation pattern of the antenna system. 

In Figure \ref{fig:aper_eff_edgetaper}, we analyze aperture efficiency by varying the antenna array size and tapering.  
A small antenna array, $<0.5F\lambda$, would not be very directive, resulting in low spillover efficiency. 
As the number of antenna elements increases, the directivity increases, and the spillover efficiency increases. Large antenna arrays $>2F\lambda$ will be very directive and approach no spillover, at the cost of being physically larger.

Furthermore, a Gaussian edge taper reduces side-lobe levels, enhancing peak spillover efficiencies. 
However, aggressive edge-tapering requires larger pixels, which conflicts with dense packing of pixels. 

Following the prescription of \cite{Griffin_2002} Section 4.2, we can compare the relative mapping speed of different detector focal planes for an extended source. The power as seen by a detector in the Rayleigh-Jeans limit has the form, 
\begin{equation}
P = \eta_0 \eta_s A \Omega \left( \epsilon \frac{k_b T}{\lambda^2} \right) \Delta \nu,
\end{equation}
where $\eta_0$ is the detector optical efficiency, $\eta_s$ is the spillover efficiency, $A\Omega$ is the detector throughput, $\epsilon$ is the emissivity of a source at temperature $T$ and $\Delta \nu$ is the bandwidth of the detector. The fundamental noise sources include the photon noise and intrinsic detector noise,
\begin{equation} 
NEP^2 = NEP^2_\gamma + NEP^2_{\text{detector}} = NEP^2_\gamma \left(1 + \gamma^2 \right), \\
\end{equation}
where we parameterize $\gamma = \frac{NEP_{\text{detector}}}{NEP_\gamma}$ as the ratio of detector noise to photon noise. Commonly, these detectors are in the photon noise limit so $\gamma < 1$. We can then make an approximation that the bunching noise term is sub-dominant to the Poisson noise term: 
\begin{equation} 
NEP^2 \approx 2 \hbar \nu_0 P \left(1 + \gamma^2 \right).
\end{equation}
The corresponding single detector NET is then, 
\begin{equation} 
NET = \frac{NEP}{\frac{\partial P}{\partial T}}  = \frac{2 \hbar \nu_0}{ \sqrt{\eta_0 \eta_s A \Omega \left( \epsilon \frac{k_b}{\lambda^2} \right) \Delta \nu}}  \sqrt{1 +\gamma^2},
\end{equation} 
A focal plane NET allows us to compare the relative mapping speeds of two different focal planes when observing the same source. The relative instantaneous sensitives of the two focal planes is expressed,
\begin{equation} 
\frac{NET}{NET'} =  \sqrt{ \frac{( N \eta_0 \eta_s A \Omega \Delta \nu)'}{N \eta_0 \eta_s A \Omega \Delta \nu}} \sqrt{ \frac{1 + \gamma^2}{1 + \gamma'^2} }.
\end{equation}

Therefore, the relative mapping speed of a single-moded, photon-noise limited detector array depends on the number of detectors, optical efficiency, spillover efficiency, and total bandwidth. 
Consequently, optimizing an array naturally finds a balance between a dense pixel packing size, $\sim 1.5F\lambda$, with maximizing spillover efficiency and optical efficiency.
Additionally, in the limit that the photon-noise contribution comes from the bunching term, the relative mapping speed only depends on the square root of number of detectors and their bandwidth.

The calculations presented here apply directly to other single-mode optical systems like horn-coupled detectors. 
Horn-coupled antennas' edge-taper varies depending on the specific details of the horn profile, complicating cross-comparisons between these two optical architectures.  
Nevertheless, antenna arrays offer a straightforward means of controlling the degree of taper. This allows them to excel in achieving a tightly controlled illumination pattern and offer the highest pixel density. 

\bibliographystyle{aasjournal}
\bibliography{references}{}
\end{document}